\documentclass[twocolumn]{openjournal}
\usepackage{graphicx} 

\usepackage[colorlinks,linkcolor=blue,citecolor=blue,urlcolor=blue ]{hyperref}
\usepackage[utf8]{inputenc}
\usepackage{float}
\usepackage{xcolor}
\usepackage{ulem}
\usepackage[T1]{fontenc}
\usepackage{amsmath}
\usepackage{xparse}
\usepackage{physics}

\newcommand{\msolaryr} {$\rm{M_{\odot} \ yr^{-1}}$ }
\newcommand{\renaissance}{\texttt{Renaissance}}

\newcommand{\zsolar}{\rm{\ Z_{\odot}}}
\newcommand{\msolar}{\rm{M}_\odot}

\begin{document}
\title{A Heavy Seed Black Hole Mass Function at High Redshift - Prospects for LISA\vspace{-1.5cm}}
\author{Joe M. McCaffrey$^{1,2,*}$}
\author{John A. Regan$^{1,2}$}
\author{Britton D. Smith$^3$}
\author{John H. Wise$^4$}
\author{Brian W. O'Shea$^{5, 6, 7}$}
\author{Michael L. Norman$^{8,9}$}
\affiliation{$^1$Department of Physics, Maynooth University, Maynooth, Ireland}
\affiliation{$^2$Centre for Astrophysics and Space Science Maynooth, Maynooth University, Maynooth, Ireland}
\affiliation{$^3$Institute for Astronomy, University of Edinburgh, Royal Observatory, Edinburgh EH9 3HJ, UK}
\affiliation{$^4$Center for Relativistic Astrophysics, Georgia Institute of Technology, 837 State Street, Atlanta, GA 30332, USA}
\affiliation{$^5$Department of Computational Mathematics, Science, and Engineering, Michigan State University, East Lansing, MI 48824, USA}
\affiliation{$^6$Department of Physics and Astronomy, Michigan State University, East Lansing, MI 48824, USA}
\affiliation{$^7$Facility for Rare Isotope Beams, Michigan State University, East Lansing, MI 48824, USA}
\affiliation{$^8$San Diego Supercomputer Center, University of California, San Diego, 10100 Hopkins Drive, La Jolla, CA 92093, USA}
\affiliation{$^9$Center for Astrophysics and Space Sciences, University of California, San Diego, 9500 Gilman Dr, La Jolla, CA 92093, USA}
\email{$^*$email: joe.mccaffrey.2018@mumail.ie}
\begin{abstract}
    \noindent The advent of new and near-future observatories probing the earliest epochs of the Universe has opened the opportunity to investigate the formation and growth of the first massive black holes (MBHs). Additionally, the use of high resolution cosmological simulations to investigate these high-redshift environments is needed to predict the dark matter halos in which these MBH seeds will form. We use the \renaissance{} simulations to analyse the formation and growth of so-called heavy seed black holes. Other past work has investigated the formation and growth of light (black hole) seeds with \renaissance{} and found that these black holes do not grow in the environments in which they reside.
    In this work we seed MBHs, in post-processing, and track accretion onto the MBHs as well as mergers with other MBHs at high-redshift.  We show that the heavy seeds struggle to achieve high accretion rates with only the most massive black holes ($\gtrsim 10^5 \msolar$) growing at close to the Eddington limit under optimistic conditions. Despite the lack of significant growth for these early MBHs, the signals from their merger events will be sufficiently strong (SNR $\sim 10^2$) to be probed by the next generation of gravitational wave observatories, such as \textit{LISA}. We predict that \textit{LISA} will observe of the order of $10$ MBH merger events per year where the mergers occur at z $\gtrsim$ 10 or at least begin their early inspiral phase at z $\gtrsim$ 10.
    
\end{abstract}
\maketitle
\section{Introduction}
\noindent Massive Black Holes (MBHs) are a common feature in almost every galactic center in the Universe \citep[e.g.][]{doi:10.1146/annurev-astro-082708-101811, 2016ApJ...818...47S, 2016ApJ...831..134V, Fan_2023}. These central MBHs can have masses that range from millions to billions of times the mass of the Sun and exist at epochs from the present day universe back to, at least, only several hundred million years after the Big Bang \citep{Ghez_2008,Larson_2023, Goulding_2023, Bogdan_2023,Maiolino_2023}. The very existence of MBHs in the very early Universe places a severe challenge to our understanding of compact object formation, growth and evolution at early times \citep{Regan_2024}. An outstanding challenge in astrophysics and cosmology is therefore how did the first seed black holes grow to the masses we observe in galaxy centres within a Hubble time \citep[e.g.][]{Volonteri_2010, Inayoshi_2020}. \\
\indent To this end models have focused on trying to explain how MBHs can form and ultimately end up powering the quasars in galactic nuclei. There are a number of potential pathways that can lead to a MBH \citep[e.g.][]{Rees_1978}. In broad terms the pathways can be broken into
the light seed pathway and the heavy seed pathway. The light seed pathway \citep[e.g.][]{Madau_2001} relies on the remnants of Population III (Pop III) stars, the first generations of stars in the Universe. The black holes formed from these stars would have an upper limit of at most 10$^3$ M$_{\odot}$ \citep{Schnedier_2002, Hartwig_2015, Lazar_2022}. In order for these black holes to reach the masses that we observe at high-z, they would need to undergo accretion at the Eddington rate for extended periods and for the high-z quasar population, in particular, accretion at, or beyond, the Eddington rate with a duty cycle of very close to unity. This may be physically unlikely as the black hole would need a consistent supply of matter to accrete from and would need to stay at the centre of a dense accretion flow \citep{Whalen_2004, Milosavljevic_2009, Alvarez_2009}. \\
\indent The dynamics of black hole accretion has turned out to be one of the main bottlenecks to efficient accretion onto light seed black holes with numerous studies showing that light seed black holes are unable to efficiently sink to the high-density centre of galaxies \citep[e.g.][]{pfisterErraticDynamicalLife2019, Ma_2021}. While dynamics, without doubt, does provide a strong bottleneck to efficient growth, growth within more favourable environment may be possible \citep[e.g.][]{Lupi_2014, Alexander_2014, Lupi_2016, Natarajan_2020, Shi_2023, Shi_2024b, Shi_2024, gordonHungryNotHow2024, Mehta_2004}. The question is how frequently this efficient (possibly via bursts of super/hyper-Eddington) accretion occurs and what is the parameter space that allows it to occur? If the probability of growth is too low then perhaps another seed formation pathway is required. \\
\indent The heavy seed pathway emerges as an attractive scenario in this case. For the heavy seed pathway the initial mass of the black holes is expected to be in the range $10^3 - 10^5\ \msolar$. The actual formation of the heavy seed can occur via two, potentially overlapping, pathways. On the one hand
we have formation via a supermassive star (SMS) in the nucleus of the halo \citep[e.g.][]{Latif_2013c, Latif_2013d, Latif_2016a,Regan_2018b,Woods_2019, Regan_2020b}, or else a heavy seed may form via dynamical runaway within a dense stellar cluster or dense black hole cluster \citep{Devecchi_2008, Katz_2015, Rizzuto_2021, Gonzalez_2021, Fragione_2018, Fragione_2022, Antonini_2019, Mapelli_2021, Arca-Sedda_2021, reinosoFormationSupermassiveStars2023}. 
Both of these scenarios have been thoroughly investigated via analytic, semi-analytic and numerical modeling over the last decade. While convergence of models is certainly lacking it is also true that heavy seed formation has been shown to be viable and with number density values shown to be potentially compatible with the observed population of high-z MBHs \citep{Bogdan_2023}. More exotic channels for black hole formation include the formation of massive protostars that are fueled by the annihilation of dark matter particles instead of nuclear fusion \cite{DMSTARS-SPOLYAR}.\\
\indent Over the last two decades a number of channels have been explored which can in principle lead to environments capable of supporting heavy seed formation. These are i) formation of heavy seeds in haloes exposed to large Lyman-Werner (LW) fields, ii) baryonic streaming velocities leading to heavy seed formation and iii) rapid halo assembly leading to heavy seed formation. In this paper we focus exclusively on the rapid-assembly channel. Further details on the other channels can be found in the review by \cite{Inayoshi_2020}. The rapid-assembly channel potentially leads to a much greater number density of MBHs at high-z being only dependent on the assembly history of the halo up to the atomic cooling limit. Recent numerical simulations by \cite{Wise_2019}, \cite{Regan_2020}, \cite{Lupi_2021} and \cite{Latif_2022} shown that inflow rates of approximately 0.1 \msolaryr into the halo centre (i.e. down to scales of order 1 - 50 pc) is sufficient to suppress the cooling effect of $\rm{H_2}$ allowing larger Jeans masses to develop. In turn this allows the formation of massive proto-stars \citep{Regan_2020, Prole_2024}. It is this pathway we explore in this paper and from this model we quantify the expected number density of heavy seeds.  \\ 
\indent The larger mass of heavy seeds compared to light seeds gives heavy seeds distinct advantages over light seeds. The larger initial masses contribute directly to a larger sphere of influence which can be represented by the Bondi-Hoyle radius \citep{Bondi_1952}. The Bondi-Hoyle radius scales as the black hole mass squared giving the heavy seed mass black hole a significantly larger probability of accreting material. Additionally, dynamical friction scales linearly with the black hole mass again meaning that heavy seed black holes are more likely to experience efficient sinking towards higher density regions compared to lighter seeds \citep{Tremmel_2015}.  \\
\indent On the observational side it is still beyond the reach of modern telescopes to observe the seed formation epoch. Recent measurements by JWST of MBHs in very early galaxies has led some authors to conclude that these measurements are inline with expectations from a heavy seed pathway \citep{Scoggins2023, pacucci2023jwst, Natarajan_2024}. This is because the MBH population that JWST has thus far revealed at high-z does appear to reflect an over-massive population with respect to the host galaxy stellar mass, albeit selection bias may yet be at play \citep{liTipIcebergOvermassive2024}. While the weight of evidence now likely points to heavy seeds being the origin for at least some of the MBHs being detected by JWST, periods of super-Eddington accretion onto a light seed cannot be ruled out \citep[e.g.][]{Lupi_2014, Shi_2024b, Mehta_2004, Lupi_2024}. \\

\section{Methodology}
\label{sec:Methodology}
\subsection{The Renaissance Simulations}
\noindent The \texttt{Renaissance} simulations\footnote{https://rensimlab.github.io/} \citep{xuPOPULATIONIIISTARS2013a,xuHEATINGINTERGALACTICMEDIUM2014a,chen_renaissance, OShea_2015} are a suite of three zoom-in regions of cosmological N-body, hydrodynamical simulations using the adaptive mesh refinement code Enzo \citep{Enzo_2014, Enzo_2019}.
The zoom-in regions originate from a parent box with a comoving volume of 40 cMpc on the side.
The \renaissance{} simulation setup is briefly described here but we refer the reader to the previous papers for more detailed discussion. \\
\indent The goal of the \renaissance{} simulation suite was to model early galaxy formation, including Pop III and Pop II star formation and feedback. In order to accomplish this, high resolution was required to allow us to resolve (mini-)halos of masses $\sim 10^6\ \msolar$ and then track their growth through accretion and mergers with other halos. 
The adaptive mesh refinement abilities that Enzo possesses allows for the increase of resolution for select, dense regions within the simulation box. Due to the computational expense of such a strategy it could not be employed over the entire parent box (40 cMpc). Instead three regions were chosen within the parent box to represent different regions that host galaxy formation of different large-scale overdensity.
The three regions are the Rarepeak (RP) (run to a redshift of $z = 15$ with dimensions of 3.8 × 5.4 × 6.6 Mpc$^3$), Normal (run to $z = 11.6$, 6.0 × 6.0 × 6.125 Mpc$^3$) and Void (run to $z = 8$, with dimensions the same as the Normal region) regions \citep{chen_renaissance}, the names referring to their level of overdensity with respect to the cosmic mean. As the goal of this study is to constrain the number density of heavy seed black holes we choose to study only the Normal and Rarepeak regions here as the Void region has been previously found to support only very few, if any, heavy seeds \citep{reganEmergenceFirstStarfree2020}. \\
\indent The \renaissance{} simulations contain models for metal free (Population III) and metal-poor (Population II) to allow for self consistent (early)
star formation.
Crucially the simulations, however, do not contain a prescription for
black hole formation \citep[see also][]{smithGrowthBlackHoles2018}.
We will use the information that the \renaissance{} simulation snapshots supply to predict regions in which massive black hole (MBH) formation is likely to occur. We then use that same information to predict the accretion and merger rates for MBHs in the early Universe. 
The resolution for the dark matter particles in \renaissance{} is $2.4 \times 10^4 \msolar$ while the gravitational softening length is resolved down as far as 19pc.

\subsection{Seeding the Massive Black Hole Population} \label{Sec:Conditions}
\noindent To identify regions of MBH formation, we analyse each halo, from its starting point within the merger trees (built by Rockstar \citep{Behroozi_Wechsler_Wu_Busha_Klypin_Primack_2012} and consistent-trees \citep{Behroozi_2012}) and work our way forward to identify the earliest point at which each halo satisfies the conditions needed for seed formation. The filters we use to identify MBHs within haloes inside the simulation volumes are determined by the following conditions:
\begin{itemize}
    \item $T_{\text{vir}} \gtrsim \rm{Atomic \ Cooling \ Limit} \ (\sim 10^4 \textrm{ K})$
    \item Metallicity, $Z  < 10^{-3} \text{ Z}_{\odot}$
    \item Compactness, $\gamma > 0.5$
    \item Inflow rate, $\dot{M} > 0.1 \text{ M}_{\odot}/\text{yr}$ 
\end{itemize}.
We now discuss each of these criteria in turn. \\

\indent We only consider haloes that exceed the atomic cooling threshold. 
Atomic cooling haloes are characterised by reaching a virial temperature at which atomic line emission cooling becomes effective. The virial temperature at which this occurs is $T_{\text{vir}} \approx 10^4 \textrm{ K}$, where 
\begin{equation}
    T_{\text{vir}} = 0.75 \times 1800\qty(\frac{M_{\text{vir}}}{10^6\,\text{M}_\odot})^{2/3}\frac{1 + z}{21}\text{ K},
\end{equation}
\citep{fernandez2014}. 
This presents us with a mass limit that we use to identify the atomic cooling halos,
\begin{equation}
    M_{\text{crit}} = 10^{12} \qty[0.75\times1800\qty(\frac{z + 1}{21})]^{-3/2}\ \msolar
\end{equation}
 As the halos collapse, due to radiative cooling, the density increases and the opportunity for massive stars and subsequent black hole formation presents itself.
%
In the absence of metals (discussed below) and thus fragmentation, a single supermassive star can form \citep{wiseFormationMassiveBlack2019, latifRoleMagneticFields2023} and thus end its life in a higher mass MBH compared to black holes formed from the Pop III stars in the light seed pathway \citep{sassanoLightMediumweightHeavy2021}. 
\\ \indent In addition to the atomic cooling limit mass threshold, we also impose a
minimum metallicity threshold of $Z < 10^{-3} \zsolar$. We require the halo to have low metallicity as the presence of metals adds an additional and very efficient coolant lowering the Jeans mass substantially, thus causing fragmentation to occur \citep{brommFragmentationPreenrichedPrimordial2001, petersLowmetallicityStarFormation2014, sassanoLightMediumweightHeavy2021}. The fragments that will form in metal enriched gas would form Pop II stars and perhaps a proto-globular cluster depending on the environmental conditions. Our criteria of filtering out all haloes with a metallicity  $Z > 10^{-3} \zsolar$ allows for the formation of both supermassive seeds, through a near monolithic collapse, and potentially also fragmentation into massive star clusters \citep{reinosoFormationSupermassiveStars2023, schleicherOriginSupermassiveBlack2022, chonSupermassiveStarFormation2020}, allowing for multiple methods in which a heavy seed can form. In this way we are combining the dynamical runaway collapse model \citep[e.g.][]{Boekholt_Schleicher_Fellhauer_Klessen_Reinoso_Stutz_Haemmerlé_2018, Tagawa_Haiman_Kocsis_2020, Das_Schleicher_Basu_Boekholt_2021, Escala_2021, Vergara_Escala_Schleicher_Reinoso_2023,Chattopadhyay_Stegmann_Antonini_Barber_Romero-Shaw_2023}and the super-massive star model into a single pathway - the formation of a heavy seed \citep{ quinlanCollapseDenseStar1987, loebCollapsePrimordialGas1994, lodatoSupermassiveBlackHole2006, begelmanFormationSupermassiveBlack2006, lodatoMassFunctionHighredshift2007, schauerFormationDirectCollapse2017, kulkarniCriticalDarkMatter2021}. 
For the purposes of this study we are agnostic to the 
emergence of the heavy seed black hole at scales we cannot resolve. 
 \\
\indent The third criteria for every MBH seed halo
to meet concerns the distribution of the mass inside the halo.
We consider a parameter we label as `compactness', which is defined as the ratio of mass inside a sphere of half the virial radius centered on the highest density point, over the total mass of the halo. 
A compactness, $\gamma$, threshold of $\gamma$ > 0.5 is used to identify halos that are sufficiently compact for seed formation to occur. Mathematically we define $\gamma$ as:
\begin{equation}
    \gamma = \frac{M(r < 0.5R_{\text{vir}})}{M(r < R_{\text{vir}})},
\end{equation}
where $\gamma = 0.5$ for a singular isothermal sphere.

By identifying compact galaxies, as given by $\gamma$, we are selecting galaxies which are likely to provide a rich environment for both initial MBH seed formation and potentially also early growth.\\
\indent Finally, we consider only halos with a mass inflow rate exceeding 0.1 \msolaryr which is needed to generate the conditions thought necessary to allow for (super-)massive stars to form \citep{hosokawaSupermassive2012,Woods_2019, wiseFormationMassiveBlack2019, Regan_2023}, thus leading into a heavy seed. The gas mass inflow rate is directly connected to the growth rate of the halo \citep[see e.g.][]{Regan_2023} with higher growth rates leading to higher dynamical heating within the halo \citep[][]{Yoshida_2003}

The mass inflow rate, which is defined as 
\begin{align}
    \frac{\partial M}{\partial t} &= \int_\sigma \rho \vec{v}\cdot d \vec{\sigma}\\
    & =4 \pi R^2 \rho v_r
\end{align}
where $\sigma$ is the sphere of virial radius $R_{vir}$ centered on the highest density point of the halo in question, $\vec{v}$ is the velocity of the gas flowing through an infinitesimal area element on $\sigma$ \citep{volume_integral, Regan_2020}. The inflow is calculated on a sphere of 20 pc radius centered on the highest density point. We choose the highest density point as the centre due to the clumpy nature of the halo. 

\indent Once all of the heavy seed halo conditions are met, we assume the formation of a heavy seed MBH. In reality the heavy seed MBH will be preceded by phase of stellar evolution either via a supermassive star, a dense stellar cluster or some combination. In either case the lifetime of the 
stellar phase of the heavy seed is expected to be of order $10^6$ years \citep{Yungelson_VanDenHeuvel_Vink_PortegiesZwart_DeKoter_2008, Woods_2019}.   Due to the fact that the time between snapshots in \renaissance{} is on the order of a few million years, we can assume that our heavy seeds can immediately be identified as MBHs.
It is noted that because we allow heavy seeds to form in halos that reach atomic cooling limit ($\sim10^6 \msolar$, along with the other criteria for seeding), this allows for black holes to form in earlier times compared to other simulations that seed at $10^{10} \msolar$ (EAGLE \cite{EAGLE}, FABLE \cite{FABLE} and SIMBA \cite{SIMBA}). Thus, giving the black holes in this model extra time to grow in comparison to other cosmological simulation. \\
\indent Some of our choices for the seeding prescription are open to investigation (as is almost universally true for all subgrid models). In particular our choice of the Inflow Rate threshold (0.1 \msolaryr) and the compactness parameter are valid avenues for further exploration. The inflow rate threshold is the parameter that ultimately dictates the number density (see Figure \ref{fig:conditions}). The value chosen is based on the high resolution studies of \cite{Regan_2020} and \cite{Regan_2023} which show that MBH seeds can form in haloes experiencing inflow rates at or above this threshold. As such we believe our choice to be robust. 


\subsection{Assigning Masses to the Seeds}
\label{method-seedingmass}
\noindent After the halos have been identified, we need to assign a black hole mass to the seeds. Since there is no constrained initial mass function (IMF) for MBH seeds, we will instead rely on observationally derived scaling relations to build our initial black hole population.
In this work we make used of two stellar mass versus black hole mass relations, the first is using data from the local Universe \citep[][using the local AGN relation, rather than the relation used for inactive elliptical and spiral galaxies]{reinesRelationsCentralBlack2015} while the second is derived from high-z JWST galaxies \citep{pacucci2023jwst}. The second relationship suggests that central MBHs are ``over-massive'' with respect to their host galaxy stellar masses. \\
\indent We also need to consider the possibility that these massive black holes can form in halos with no or negligible stellar matter. Therefore, we derive an additional relation for the black hole mass as a function of the stellar mass for halos with a stellar mass of $<10^3 \text{M}_\odot$ directly from the \renaissance{} datasets.

For the stellar mass relations, we use the following relationship when seeding black holes in haloes with stellar masses in excess of $10^3$ M$_\odot$. 
\begin{equation} \label{Eqn:Mstar_BH_pacucci}
    \log\left(\frac{M_{\text{BH}}}{M_{\odot}}\right) = \beta + \alpha\log\left(\frac{M_*}{M_{\odot}}\right)
\end{equation}
In \cite{pacucci2023jwst} (the high redshift, overmassive case) the mean value of the parameters were constrained to be $\alpha = 1.06$ and $\beta = -2.43$. For the local relation \citep{reinesRelationsCentralBlack2015}, the mean values are $\alpha = 1.05$ and $\beta = 7.45$. 
For seeding MBHs in haloes with stellar masses less than $10^3$ M$_\odot$ we use the following relation:
\begin{equation}
    \log\left(\frac{M_{\text{BH}}}{M_{\odot}}\right) = \delta + \gamma\log\left(\frac{M_{\text{vir}}}{M_{\odot}}\right),
    \label{eq:relation}
\end{equation}
 For the \cite{pacucci2023jwst} relation, the free parameters are set to be $\gamma = 1.9$ and $\delta = -11.9$. While for the \cite{reinesRelationsCentralBlack2015} relation, the  free parameters are set to $\gamma = 1.9$ and $\delta = -13.78$ respectively. We hereafter refer to the seeding method using the relation from \cite{reinesRelationsCentralBlack2015} as RV15 and to the method using \cite{pacucci2023jwst} as P23. Finally, we impose a minimum black hole seeding mass of M$_{\rm{BH}}$ > 100 M$_\odot$. We include black holes with masses less than $10^3 \msolar$ to investigate whether light seeds in lower mass galaxies are able to grow enough in order to reach masses comparable to heavy seed black holes.
 We decided to use a range of black hole seed masses, in contrast to other simulations (e.g. EAGLE \cite{EAGLE}, FABLE \cite{FABLE} and SIMBA \cite{SIMBA}). Using a range of masses will allow us to study the impacts black hole mass will have on the accretion calculations onto the black holes as well as the kind of mass ratios we expect to see in black hole mergers.

\subsection{Growth of the MBH population} \label{Sec:MBHGrowth}
We allow the seeded MBHs to grow under accretion from the surrounding gas reservoirs - as derived from the \renaissance{} datasets. To model the growth through accretion we follow the method of \cite{Smith_2018} where they grow stellar mass black holes in post-processing using the Bondi-Hoyle method \citep{1939PCPS...35..405H}, 
\begin{equation}
    \dot{m}_{\text{B-H}} = \frac{\alpha \pi G^2 M_{\text{BH}}^2\rho}{\textrm{max}\left(|\vec{v}|, c_\text{s}\right)^3},
    \label{accretion}
\end{equation}
where the boost factor $\alpha$ is included to account for the fact that the Renaissance simulations cannot fully resolve the Bondi-Hoyle radius, given by
\begin{equation}
    r_b = \frac{2GM_{\text{BH}}}{c_\text{s}^2}
    \label{bondi-radius}
\end{equation}

The choice for the value of $\alpha$ is discussed later in \S\ref{sec:boost}. The mass of the MBH at each timestep is given by
\begin{equation}
    M_{i + 1} = M_i + \dot{m}_{\text{B-H}}\Delta t
\end{equation}



\begin{figure}[htbp]
    \centering
    \includegraphics[width = 0.45\textwidth]{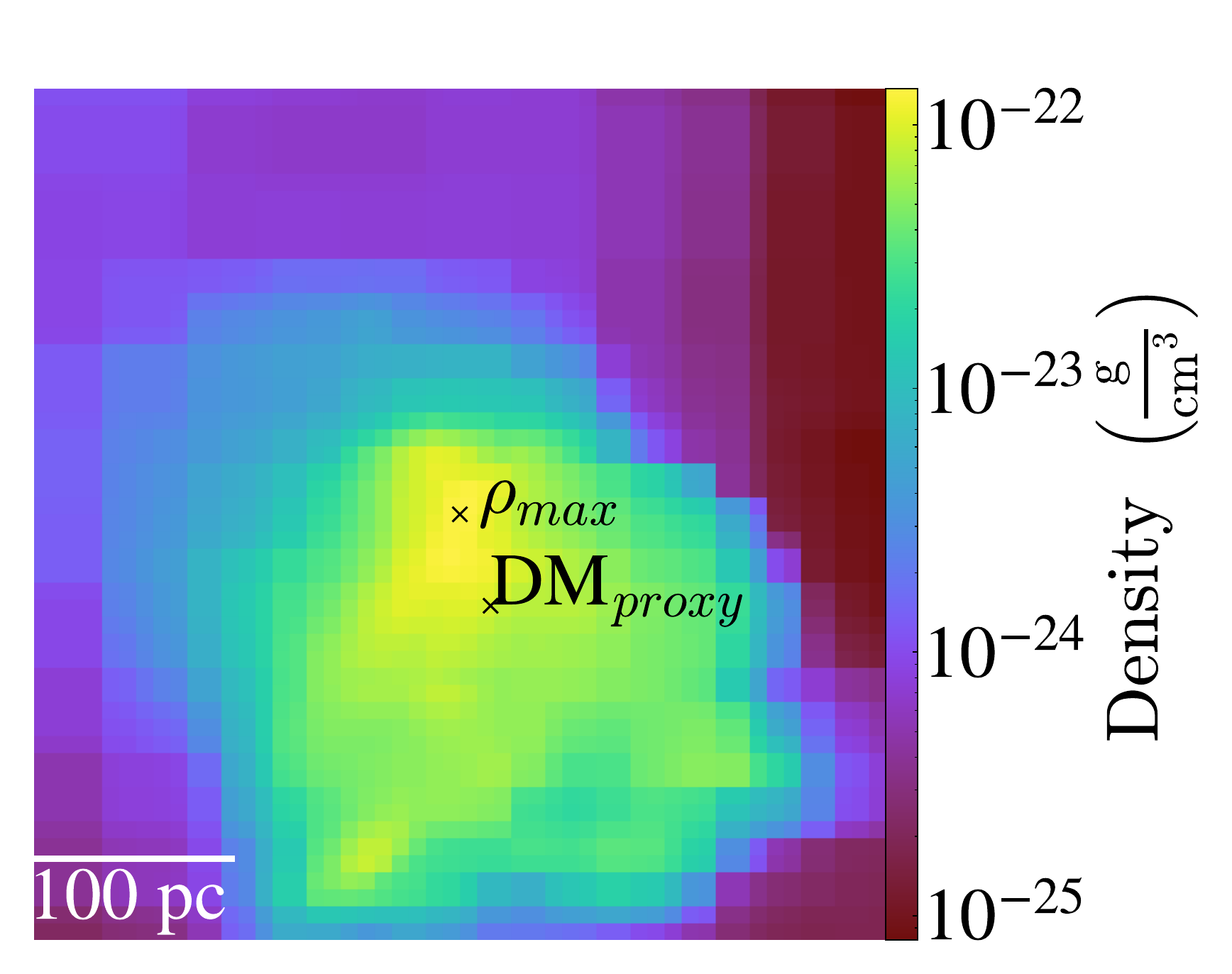}
    \caption{An example of the two methods by which the accretion onto the black holes are calculated. $\rho_{max}$ represents the position of highest density in the halo, the most optimistic case for the accretion. While DM$_{\rm proxy}$ represents the position of the black hole proxy particle (a dark matter particle from the simulation), a more realistic situation for the black hole, taking into account the wandering nature of a black hole around its host halo. The stellar mass of this halo is $7\times 10^5\  \msolar$. The MBH in this galaxy is of the order of $10^5\ \msolar$, using the overmassive \cite{pacucci2023jwst} relationship. }
    \label{fig:BH-positions}
\end{figure}

The next question is where to position our seeded MBHs at each timestep (i.e. at every output dataset). For this we take two different approaches. In the first case we simply place the MBH at the point of maximum (gas) density in each host halo and keep it there at each timestep, the $\rho_{max}$ method. The highest density point is recorded independently of the black holes existence via the halo merger tree, so there is no chance of the black hole moving to a nearby galaxy that contains a more dense region than the original halo. Taking the highest density point as the default position of the MBH is clearly a somewhat optimistic approach in this model. Realistically, however, this method may not be the most optimistic approach, due to feedback effects from star formation in the dense local environment. This stellar feedback may effectively expel a substantial amount of the gas from the local reservoir, that would otherwise be accreted by the massive black hole still in the dense environment \citep{duboisBlackHoleEvolution2015}. There is also the matter of feedback from the black hole itself during its accretion cycle, which could also impact the surrounding gas density, thus leading to slight inaccuracies in the accretion calculation, see \cite{sijackiIllustrisSimulationEvolving2015, schayeEAGLEProjectSimulating2015, bennettGrowthGargantuanBlack2023, niUltramassiveBlackHoles2022, trebitschAstraeusVIHierarchical2022} Due to the relatively low mass of even the most massive galaxies in Renaissance (M$_{*, max} \sim 5 \times 10^7 \ \msolar$) the central nucleus of the halo will be poorly defined and the central potential relatively shallow. Moreover, the mass of the seed black hole (M$_{\rm{seed}} \lesssim 10^4 \text{ M}_{\odot}$) does not guarantee that it will sink to the center of the halo. In fact, it has been shown from simulations that MBHs with masses below $10^4 \text{M}_{\odot}$ do not sink to the center within a Hubble time \citep[e.g.][]{pfisterErraticDynamicalLife2019, Ma_2021}. This is therefore the most optimistic and simplest approach and almost certainly not wholly realistic. To counter this we also take a second approach. \\
\indent The \renaissance{} simulations, like other cosmological N-body simulations,  consist of dark matter particles which make up the host halos of galaxies. 
To gain a more realistic view of the accretion onto the MBH, we also use the nearest dark matter particle to the highest density point of the halo, at the time of seeding, as a MBH proxy (the DM$_{\rm proxy}$ method). We then track this particle, via its unique particle ID as it travels within the central regions of the galaxy. In this way we use the DM particle and in particular its dynamics to create a more realistic position for the MBH seed. An example for the positions of the black holes, corresponding to each accretion method is shown in Figure \ref{fig:BH-positions}. \\
\indent Over all of our DM$_{\rm proxy}$ positions and $\rho_{max}$ positions the median distance from both the DM$_{\rm proxy}$ and $\rho_{max}$ positions to the DM centre of mass, as calculated by Rockstar, is approximately $10^2$ pc. The seeding using both methods is therefore initially consistent with the centre of mass approach, as used by many subgrid approximations.

\par
Finally, due to the fact that we post-process the accretion rates via the \renaissance{} snapshot data, we assume that the accretion is constant between snapshots. In reality, due to the clumpy nature of the galaxy, the black hole will experience irregular periods of accretion. Our methodology therefore cannot account for the likely episodic nature of accretion between snapshots and is instead a time-averaged rate. The interval between snapshots is $\Delta z = 0.1$.

\subsection{Boosting the Accretion}
\label{sec:boost}
\begin{figure*}
    \centering
    \includegraphics[width=0.45\textwidth]{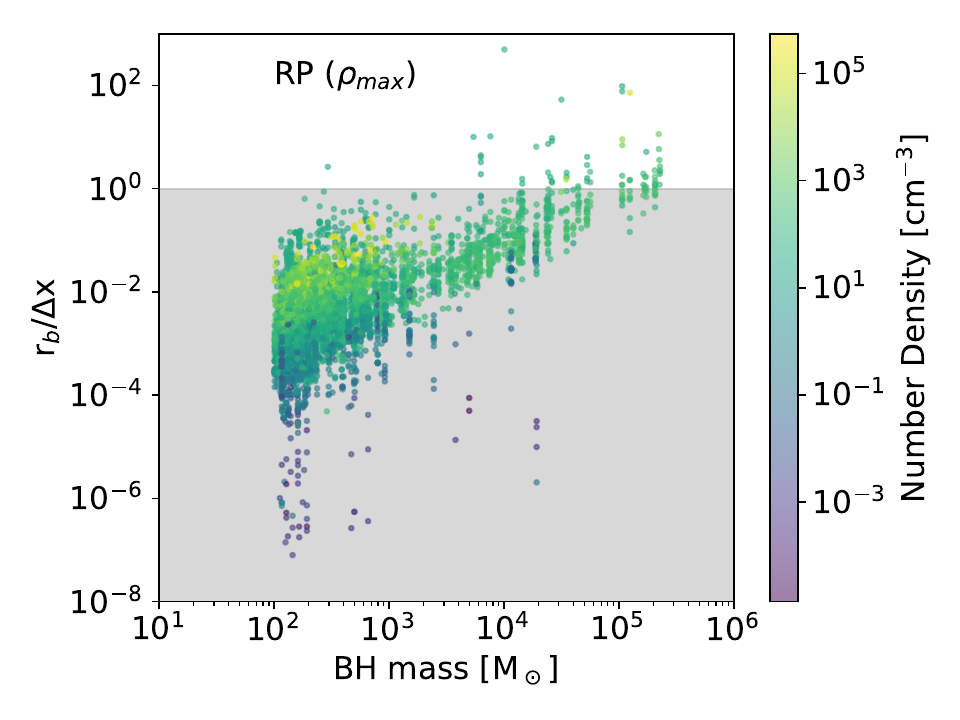}
    \includegraphics[width=0.45\textwidth]{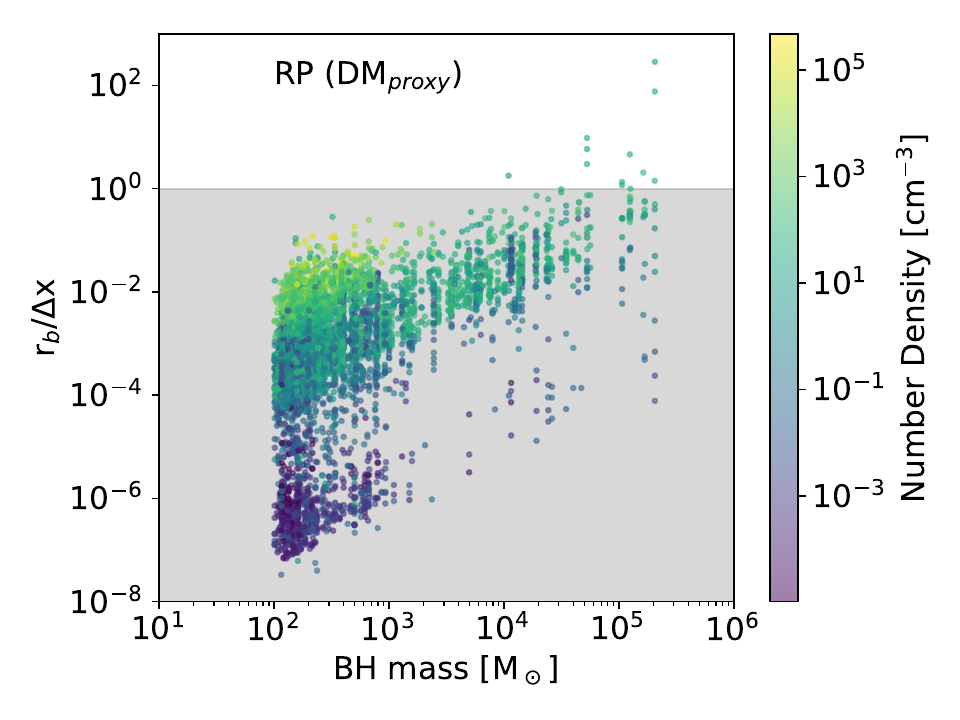}\\
    \includegraphics[width=0.45\textwidth]{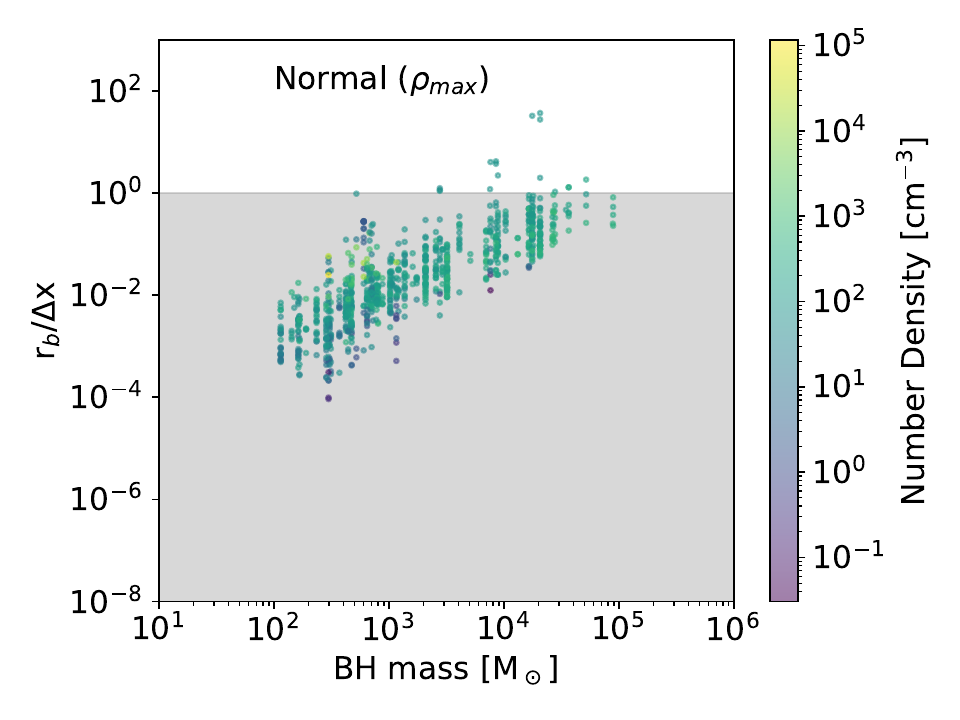}
    \includegraphics[width=0.45\textwidth]{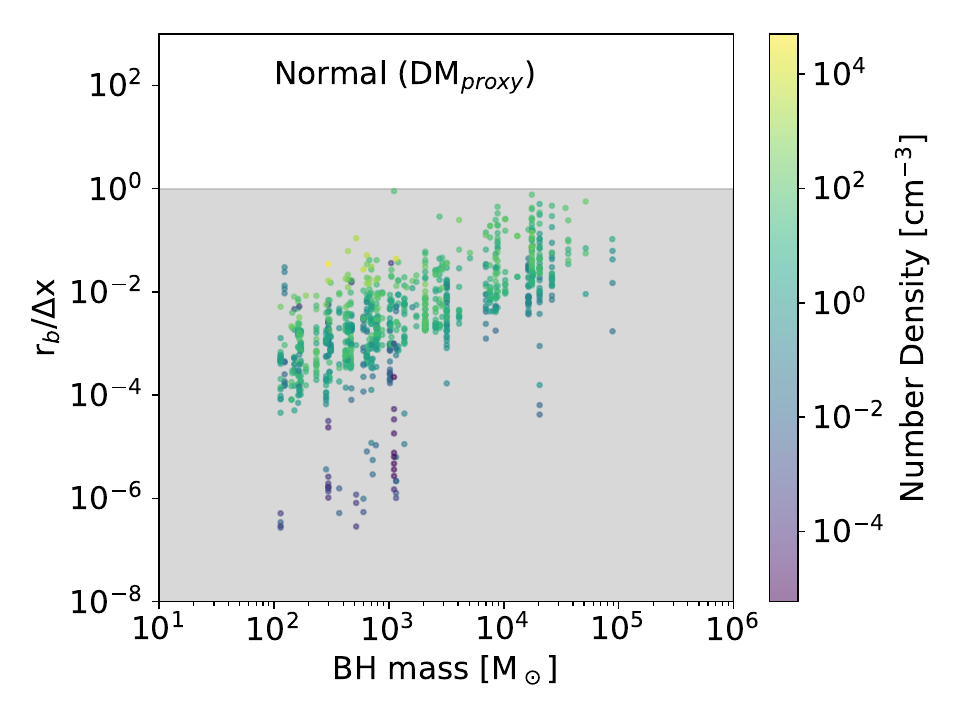}
    \caption{The population of Massive Black Holes showing the ratio between the Bondi radius, $r_b$, and the cell length, $\Delta x$, for the two black hole positions, for each region from \renaissance{}. The masses for the black holes are from P23. Using the RV15 method would only make the Bondi radii less resolved, thus further requiring the boost factor for the accretion calculation. The z-axis represents the number density of gas in the region surrounding the black hole. The grey region represents the black holes where the Bondi radius is unresolved.}
    \label{fig:bondi_radius}
\end{figure*}

As can be seen from eq.\eqref{accretion}, we include a boost factor, $\alpha$, to account for the fact that \renaissance{} may underestimate the accretion onto a MBH, in unresolved regions where the cell length is larger than the Bondi radius described by eq.\eqref{bondi-radius}. Only MBHs where the Bondi radius is not resolved are boosted.
The population of black holes where the Bondi radius is unresolved/resolved is seen in Figure \ref{fig:bondi_radius}.
The use of the boost factor was first described in \cite{springelModellingFeedbackStars2005}, however instead of using a constant value for $\alpha$, we let it depend on parameters relating to the environment around the MBH.
\par
In this case we adopt the prescription from \cite{boothCosmologicalSimulationsGrowth2009}, where they use the surrounding number density of gas to calculate the boost factor, more precisely,
\begin{equation}
    \alpha = \begin{cases}
        \left(\dfrac{n_\text{H}}{n_\text{H$^*$}}\right)^\beta & \text{for $n_\text{H} > n_\text{H}^*$}\\
        1 & \text{otherwise}
    \end{cases}
    \label{eq:boost}
\end{equation}
where the parameter $\beta = 2$. While \cite{boothCosmologicalSimulationsGrowth2009} use a critical density of $n_{\text{H}}^* = 0.1 \text{ cm}^{-3}$, we adopt the more appropriate value of $n_{\text{H}}^* = 10^4 \text{ cm}^{-3}$, for our case where the chemistry is somewhat different and the gas is able to cool to much lower values and hence the density increases. At this density, the H$_2$ starts to rapidly increase with density, the chemistry starts to change and gas starts to cool rapidly - a phase change which \renaissance{} largely misses.
\par
We adopt the boost factor implementation only for the DM$_{\rm proxy}$ method for our calculations. The $\rho_{\text{max}}$ method is already an optimistic case for the accretion onto the black holes and we believe that adding a boost for this model will only make this case even less physically meaningful.

\subsection{Tracking Mergers of Massive Black Holes}
\noindent To track the MBHs as they merge and grow with time, we need to track the progress of the host dark matter halos.
To do this \texttt{Rockstar} \citep{Behroozi_2012} is used to build a halo catalog which contains information on the growth and merging of halos with decreasing redshift. Consistent-trees was used to make a merger catalog from the Rockstar data \citep{Behroozi_Wechsler_Wu_Busha_Klypin_Primack_2012}.
We then analyse the consistent-trees outputs using \texttt{ytree} \citep{ytree}. As a halo merges with another halo we employ two different methods to calculate the merger timescale in the case where the two haloes contain MBHs. On the one hand we assume instantaneous MBH mergers (which gives the most optimistic merger times) and in the other we assign delays to the MBHs. We now discuss the implementation of delay times in our model. 
 

\subsection{MBH Merger Delay Times} \label{Sec:DelayTimes}
The four stages of black hole mergers are: dynamical friction \citep{chandrasekharDynamicalFrictionGeneral1943}, stellar loss cone \citep{mikkolaEvolutionBinariesField1992, quinlanDynamicalEvolutionMassive1996, sesanaInteractionMassiveBlack2006}, circumbinary discs \citep{heathOrbitalEvolutionBinaries2020, franchiniCircumbinaryDiscSelfgravity2021} and energy loss due to gravitational wave emission \citep{petersGravitationalRadiationPoint1963}. We can therefore add in a delay time due to an analytical prescription for dynamical friction. We do not calculate the time delays for the final three stages. To do so we would need information on the stellar populations and eccentricity of the black hole binaries. Neither of these quantities are available to us. Since we only calculate the delay due to dynamical friction the time delays we impose should be taken as lower limits.\\
\indent Despite the clumpy nature of the high redshift galaxies, the dynamical friction from the MBHs passage through the interstellar matter of the host halo can be approximated as an isothermal sphere where the time delay due to the dynamical friction is given by \citep{binneyGalacticDynamicsSecond2008, chakrabortyProbingSixMassive2023}
\begin{equation}
    t_{\text{df}} = 0.67 \qty(\frac{a}{4 \text{ kpc}})^2 \qty(\frac{\sigma}{100 \text{ km s}^{-1}})\qty(\frac{10^8\ \msolar}{M_{\text{BH},2}})\frac{1}{\Lambda} \text{ Gyr}
\end{equation}
where $a$ represents the initial distance between the two black holes. The stellar velocity dispersion, $\sigma$, is given by 
\begin{equation}
    \sigma = \qty(0.25\frac{GM_*}{R_\text{eff}})^{1/2}
\end{equation}
where $M_*$ is the stellar mass of the post merger galaxy and $R_\text{eff} = 0.1R_\text{vir}$.
Finally, the Coulomb logarithm is defined as 
\begin{equation}
    \Lambda = \ln(1 + \frac{M_*}{M_{\text{BH}, 2}})
\end{equation}
We also assume that the stellar mass of the halo remains constant as the black holes move through it. In reality and depending on the time scale of the MBH binary merger this may not be the case. Nonetheless, 
the approximation gives us a sense of time scale for the binary merger and particularly whether or not we would expect the binary to merge prior to z = 0. 


\section{Results}
\label{sec:results}
The premise of this work is to postprocess the data from the \renaissance{} datasets in order to first of 
all identify heavy seed haloes and then within those haloes to examine the growth of heavy seeds
via accretion and mergers. We therefore begin by identifying sites for heavy seed formation. 
\subsection{MBH Host Halo Identification}
\begin{figure}[tbp]
    \centering
    \includegraphics[width = 0.5\textwidth]{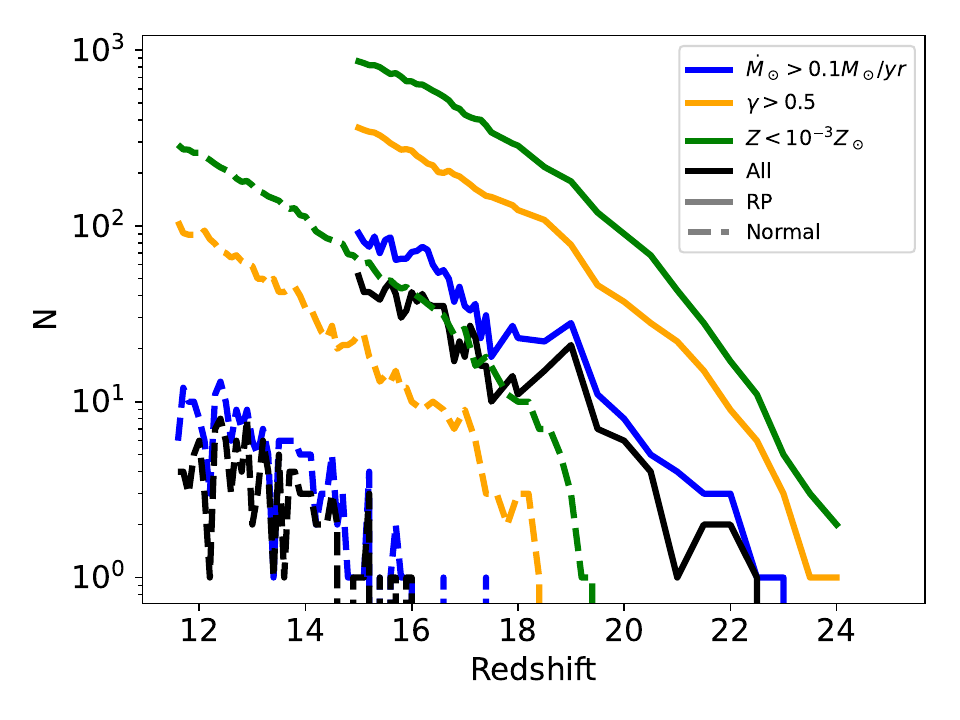}
    \caption{The number of halos that satisfy each and all of the conditions for a heavy black hole seed host halo. solid line represents the RP region, while the dashed line represents the Normal region. Once a halo meets all the seeding conditions, the halo is seeded if the corresponding MBH mass is greater than $10^2 \text{ M}_\odot$. As we can see from the two panels the mass inflow rate criteria is the strictest with the other criteria (metallicity and compactness) only having second order effects. }
    \label{fig:conditions}
\end{figure}

Using the metadata from the Normal and RP regions from \renaissance{}, we are able to search for halos which satisfy the metallicity conditions, inflow rate conditions and compactness conditions. In Figure \ref{fig:conditions} we show the number of halos that satisfy each and all of these conditions. The solid lines in  Figure \ref{fig:conditions} are for the RP region, the dashed lines for the Normal region. The strictest condition for the identifying haloes is the mass inflow criteria. This is the criteria which dominates the seeding identification. In each case only haloes with masses in excess of the atomic cooling limit are considered (condition 1, see \S \ref{Sec:Conditions}). Following this, the metallicity condition (Z $< 10^{-3} \zsolar$) is the easiest condition to satisfy (green line). The next ``easiest'' condition to satisfy is the compactness condition followed by the inflow rate condition. The black line in each panel gives the total number of haloes satisfying our conditions for heavy seed production at each timestep. 

Once a halo is seeded with a black hole, it cannot be seeded again. 
From Figure \ref{fig:conditions} it is clear that for overdense regions such as the RP region the number of heavy seeds forming is greater overall and that the haloes are also seeded much earlier at $z\sim 22$, in contrast to the Normal region which starts seeding at $z\sim 16$. Because of the criteria of the halo to reach atomic cooling limit ($\sim10^6 \msolar$), and the capabilities of \renaissance{} to resolve halos of this mass, we can be assured that \renaissance{} is able to capture the possible halos that are seeding black holes at high redshifts. Out of all the halos identified in the RP region using Rockstar, 371 ($\sim$ 4 \% of atomic cooling halos) are seeded with central MBHs (by z = 15). For the Normal region, 53 ($\sim$ 1\%) are seeded with central MBHs (by z = 11.6). We note that the conditions required for seeding show no signs of slowing down after the cessation of each region, leaving the possibility of additional black holes being formed in these high redshift regimes.\\
\indent It is clear from Figure \ref{fig:conditions} that the level of isolation that permits black holes from seeding in \renaissance{} plays an important role. We see that an overdense region, such as RP, is able to produce black holes earlier and at an order of magnitude higher at its simulation end time compared to that of the Normal region. In contrast, the Void region, which exists at a higher level of isolation fails to produce any halos capable of seeding heavy seed black holes.

\begin{figure}
    \centering
    \includegraphics[width=0.45\textwidth]{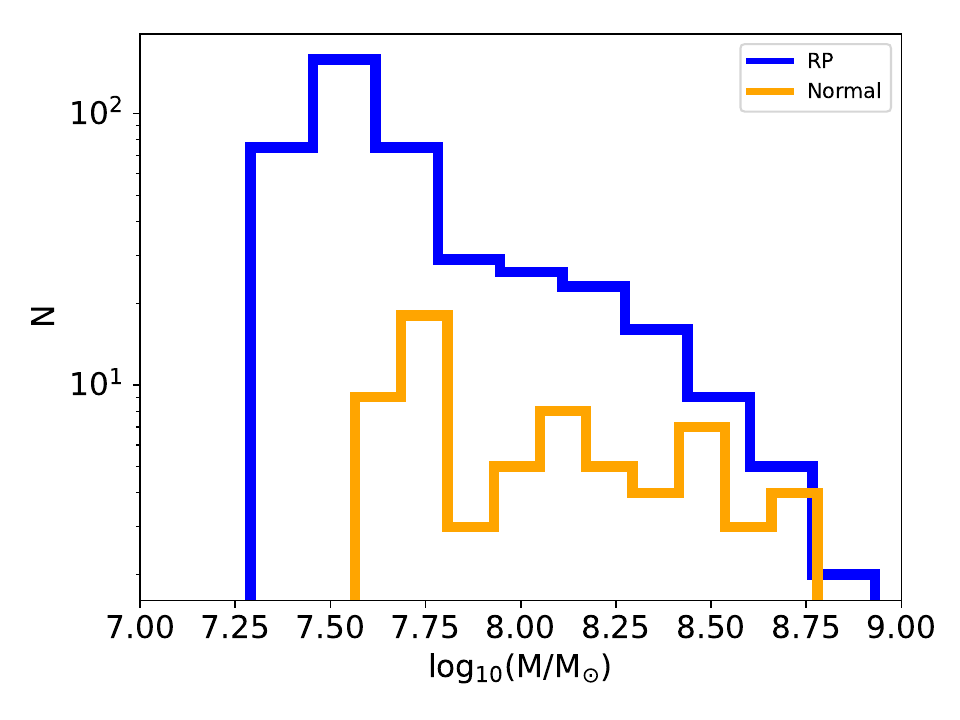}
    \caption{Histogram showing the distribution of virial masses of the host halos at the time of their respective black hole seeding.}
    \label{fig:host-imf}
\end{figure}

\begin{figure}[htbp]
    \centering
    \includegraphics[width = 0.45\textwidth]{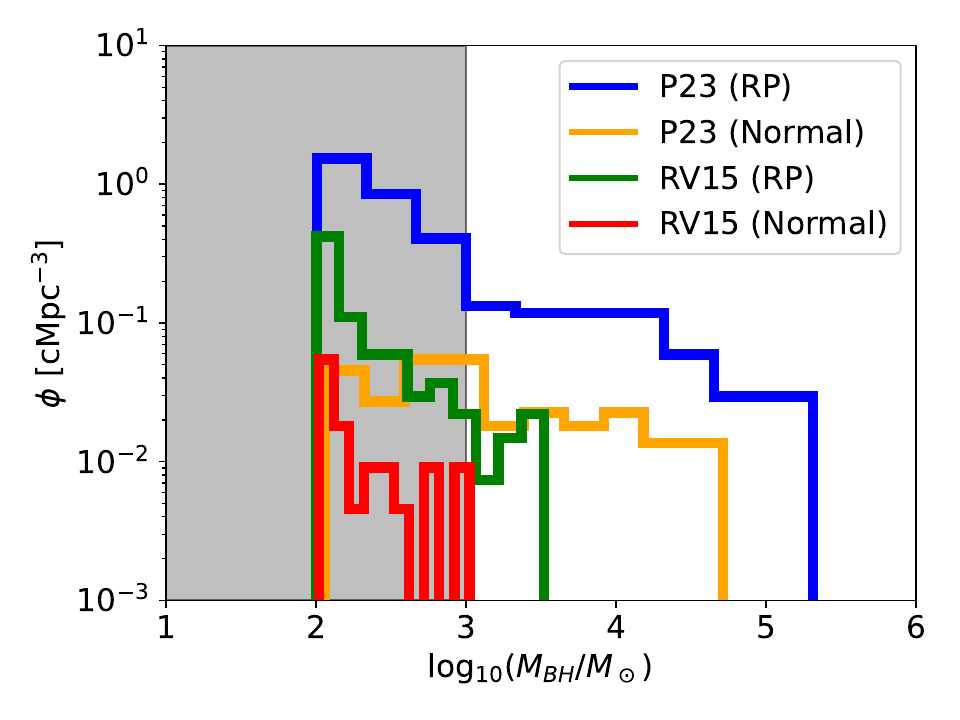}
    \caption{Initial Mass Function (IMF) of the MBHs that we seed in post-processing of the Renaissance Simulations once all the conditions for seeding in the halo have been met. We show here the IMF for both the Rarepeak region and the Normal region for both $M_{\text{BH}} -M_*$ scaling relations that we employ, from \cite{reinesRelationsCentralBlack2015} and \cite{pacucci2023jwst}. The grey shaded region denote BH masses $<10^3\ \msolar$ ie. light seeds. Everything above $10^3\ \msolar$ we consider as a heavy seed.}
    \label{fig:IMF}
\end{figure}

\begin{figure*}[tbp]

\centering

\begin{minipage}[b]{0.4\linewidth}
    \includegraphics[scale = 0.5]{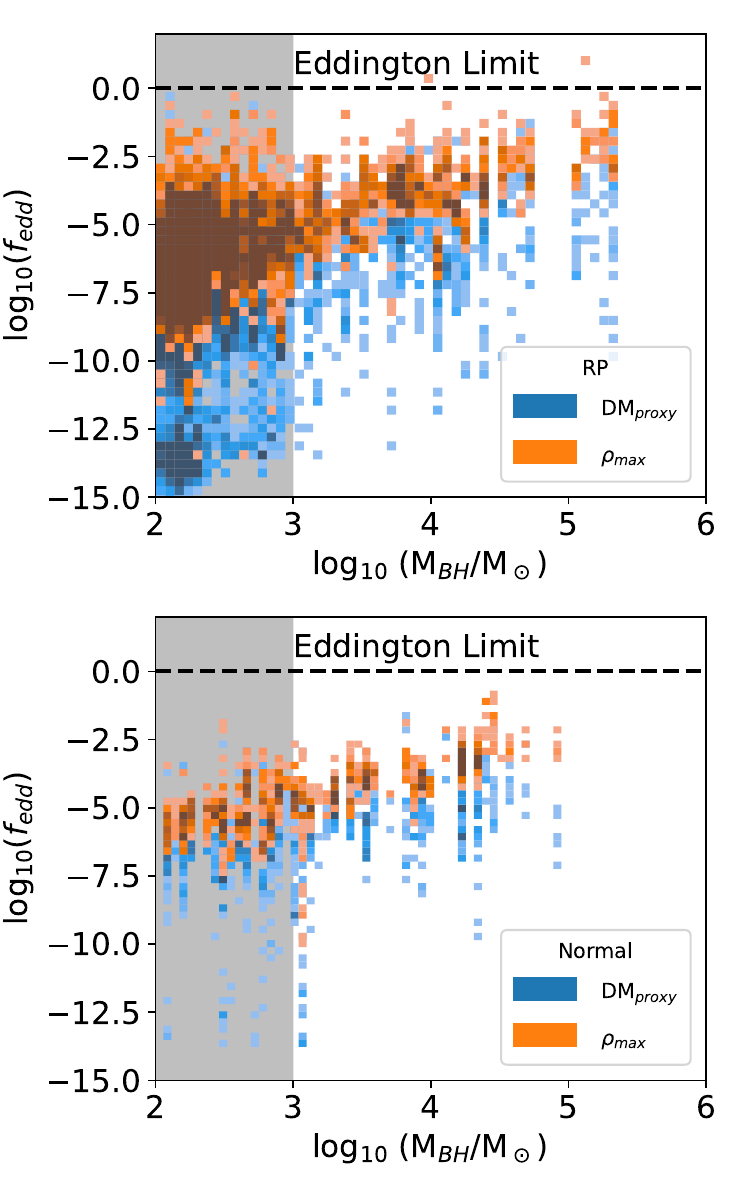}
\end{minipage}
\qquad
\begin{minipage}[b]{0.4\linewidth}
\includegraphics[scale = 0.5]{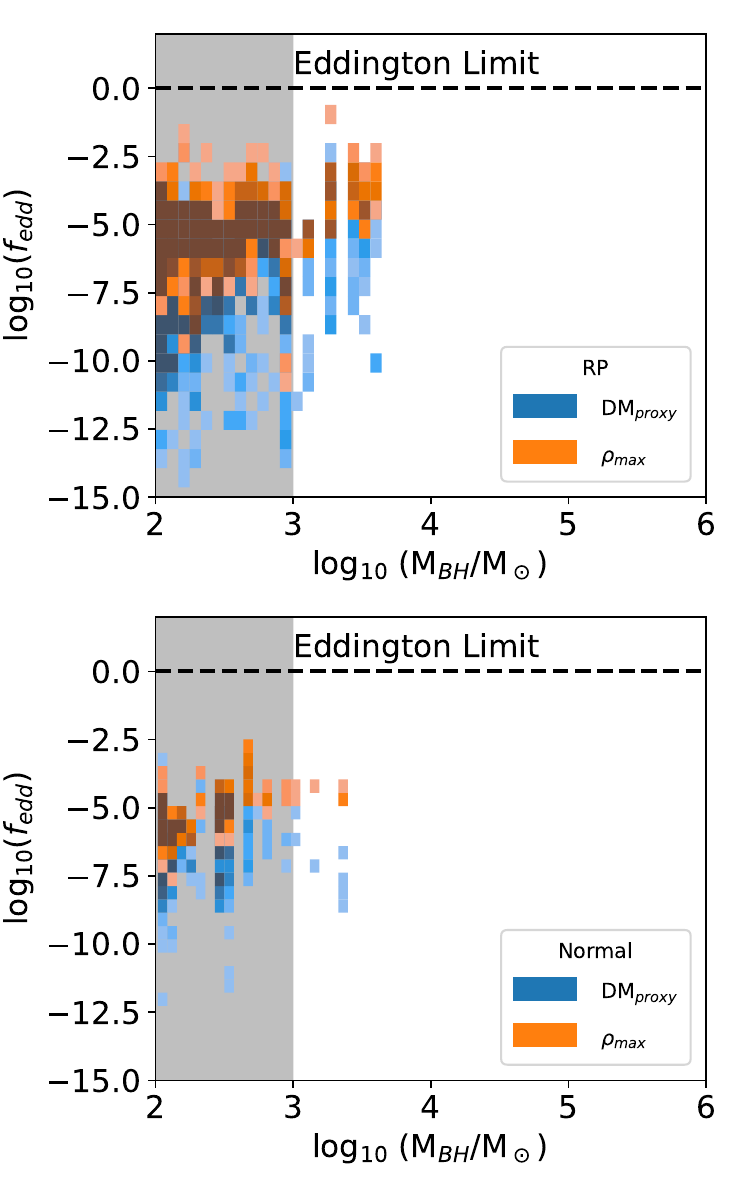}    
\end{minipage}
    \caption{2-D histograms of the Eddington fraction, for the massive black holes that exist in all the redshift snapshots from each region in \renaissance{}, in terms of Eddington fractions of the massive black holes from the Rarepeak region (top) and the Normal region (bottom). The orange represents the accretion values calculated via the $\rho_{\text{max}}$ method, while the blue represents the accretion via the DM$_{\rm proxy}$ method. The left column represents the black holes that are seeded from the P23 scaling relation, while the right is for the black holes seeded with the RV15 relation. A dashed black line showing the Eddington limit is displayed at the top of each of the figures for reference. The grey shaded regions denote BH masses $<10^3\ \msolar$.}
    \label{fig:edd_frac}
\end{figure*}

\begin{figure*}
    \centering
    \includegraphics[width = 0.3\textwidth]{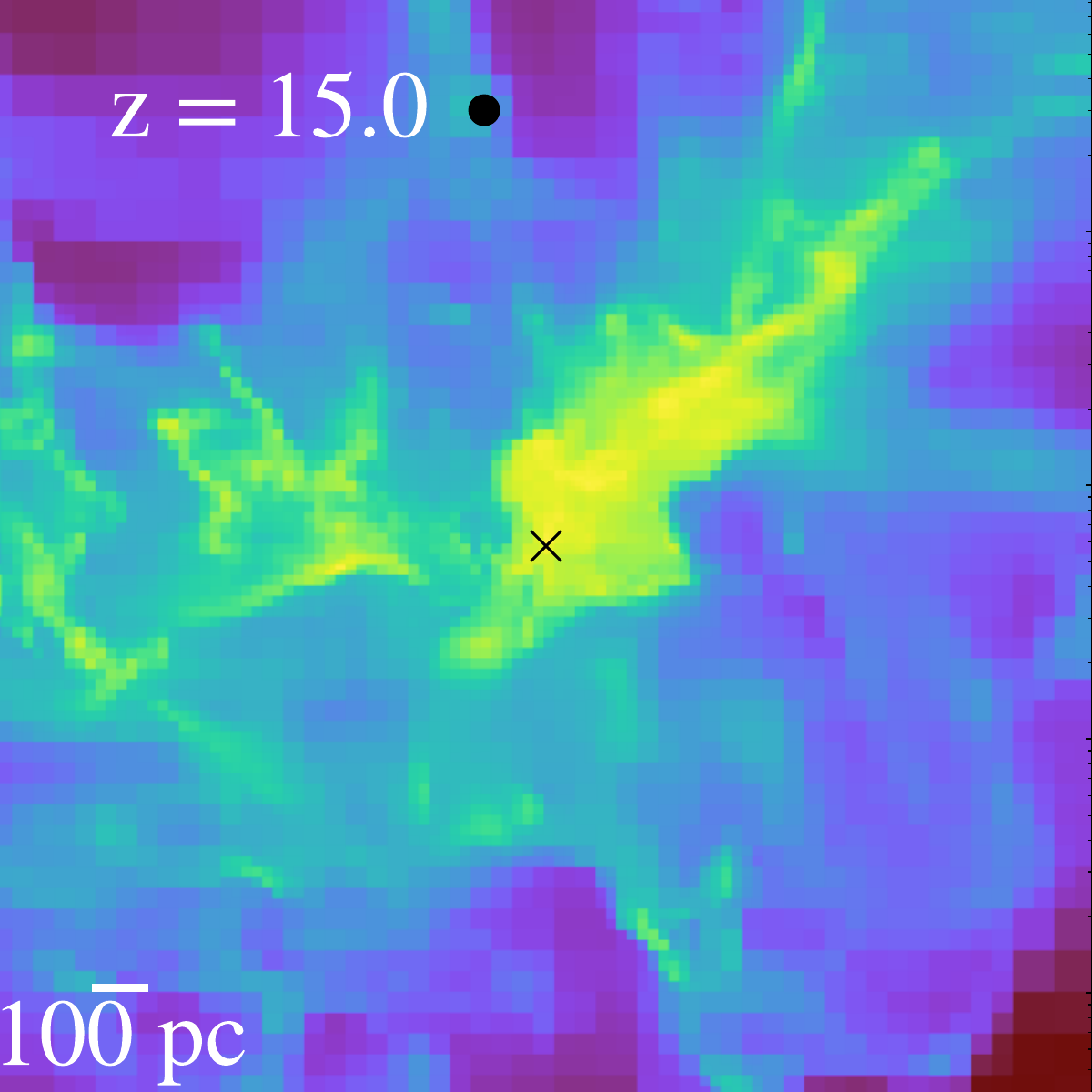}
    \includegraphics[width = 0.3\textwidth]{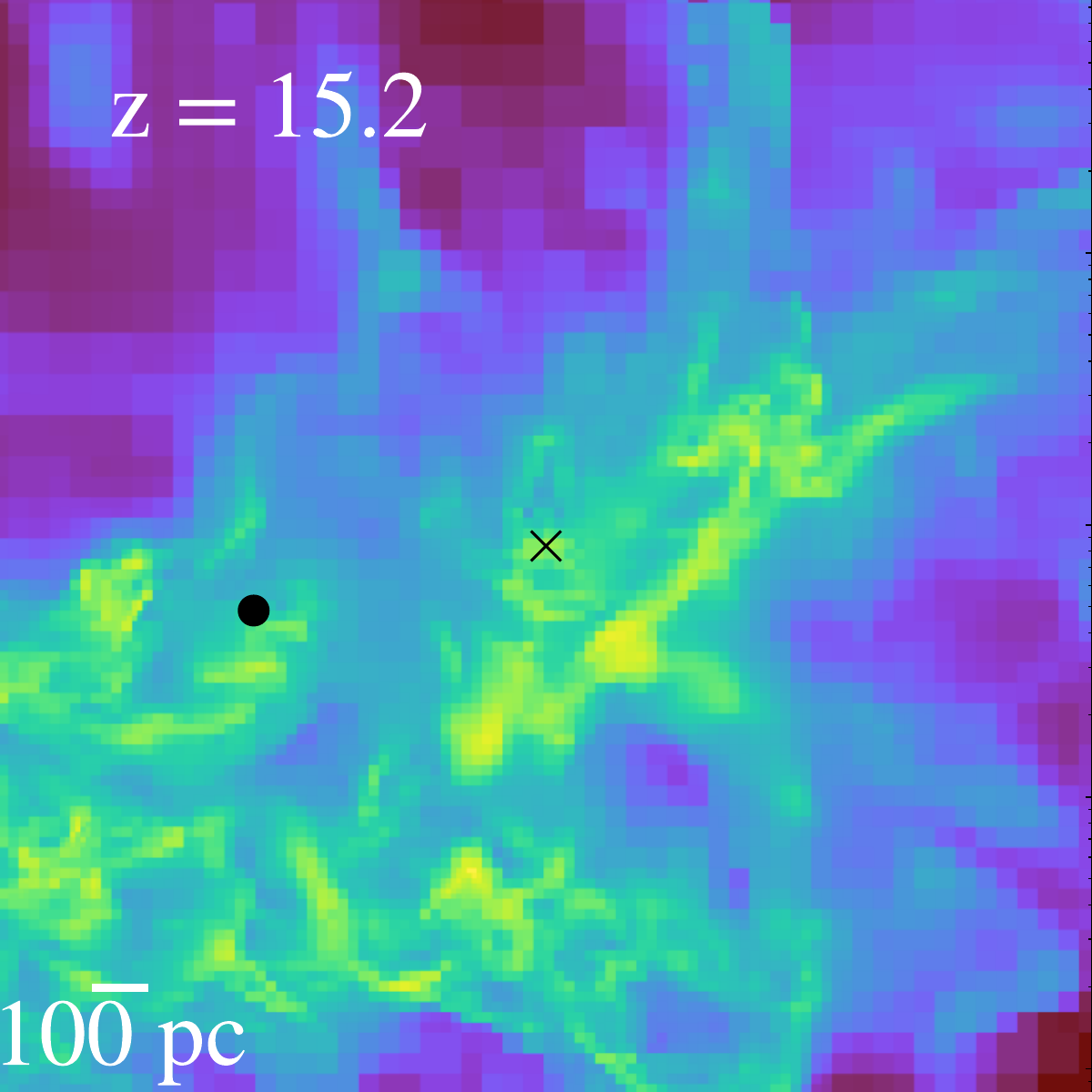}
    \includegraphics[width = 0.3\textwidth]{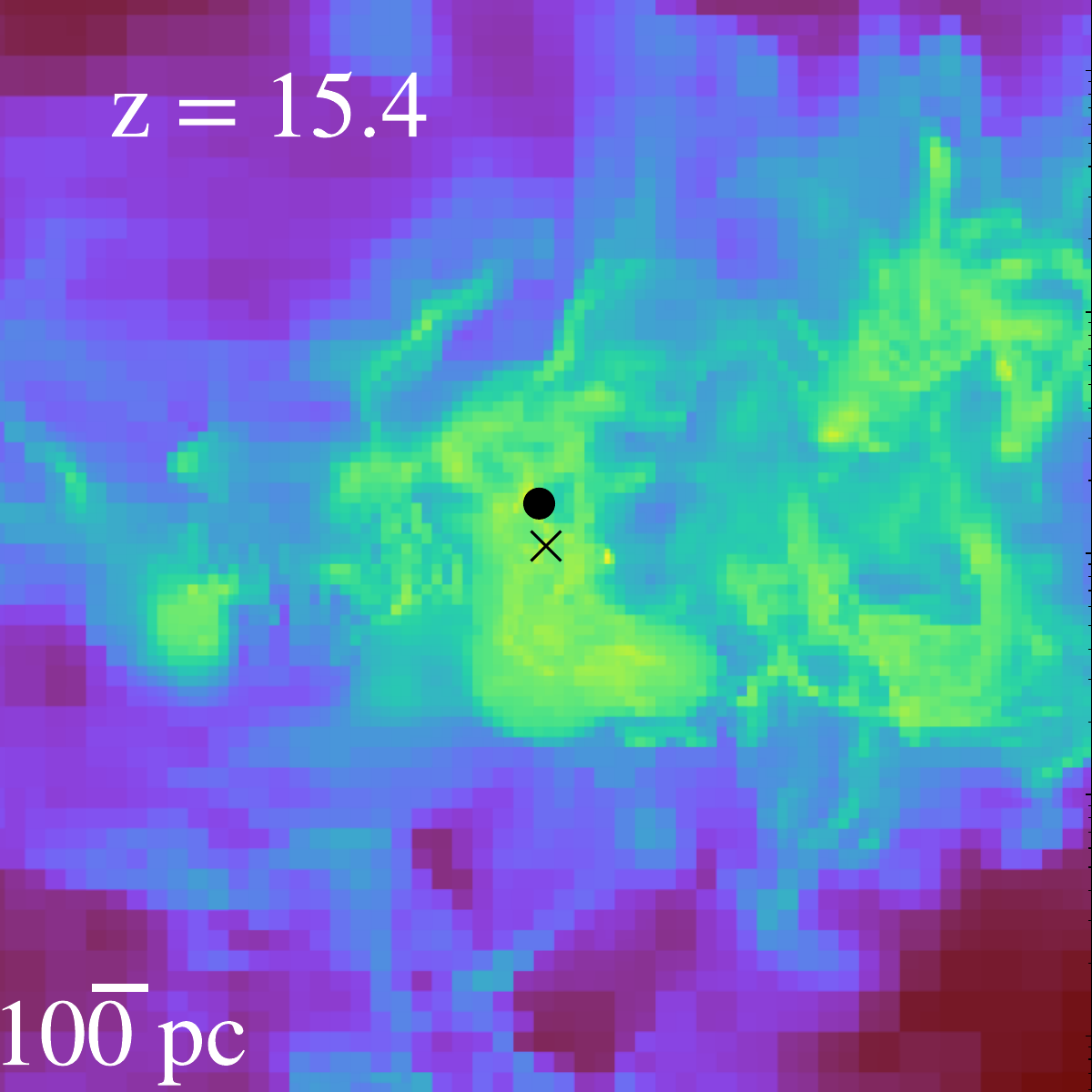}
    \caption{A series of examples showing the wandering nature of the DM$_{\text{proxy}}$ BH particle. The dot represents the proxy particle while the cross represents the region of highest density in the host halo. This example was taken from the Rarepeak region with the host halo reaching an order of $2\times 10^9 \msolar$.}
\end{figure*}

\begin{figure}
    \centering
    \includegraphics[width = 0.45\textwidth]{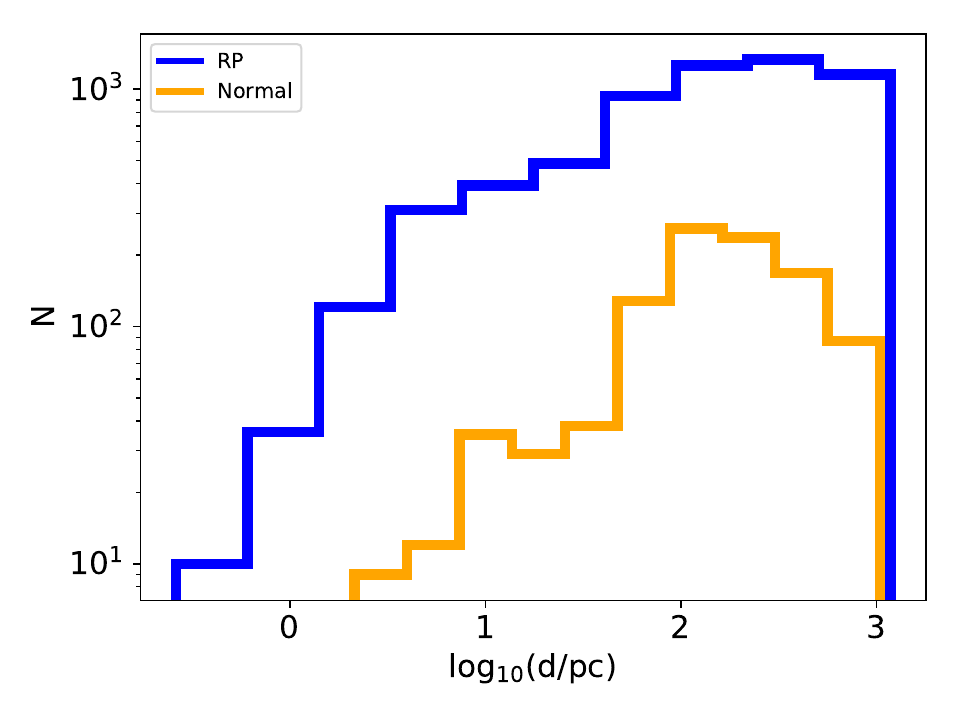}
    \caption{Histogram showing the all the distances between the DM$_{\text{proxy}}$ particle with its respective $\rho_{\rm max}$ position, at all redshifts outputs, for both the RP and Normal region.}
\end{figure}

\subsection{The MBH Mass Spectrum from Renaissance Galaxies}

Once the halos that satisfy the heavy seeding conditions have been identified, the scaling relations, that were discussed in section \ref{method-seedingmass}, are used to assign seed masses to black holes. As discussed, we employ two separate scaling relations, one derived from the high-z ``over-massive'' black hole population observed primarily by JWST \citep{pacucci2023jwst} and the other derived from the local MBH-Galaxy Mass relationship \citep{reinesRelationsCentralBlack2015}. As a result we end up with two (slightly) different black hole populations that we analyse independently.  In Figure \ref{fig:IMF} we show the spectrum of initial MBH masses than emerge from employing our scaling relations.  After seeding and assigning masses to black holes in halos that meet the heavy seed conditions, we allow them to grow within their environment. The accretion onto each MBH is then calculated using the Bondi-Hoyle formula as described in \S\ref{Sec:MBHGrowth}.

\subsection{The Growth of Heavy Seed MBHs in Renaissance}
\noindent Accretion onto the MBHs is calculated using the analytical Bondi-Hoyle formula (\S\ref{Sec:MBHGrowth}) for two different values for the density, $\rho$: one value representing the highest density point of the halo ($\rho_{max}$), and the other value extracted from the position of the dark matter proxy particle (DM$_{\rm proxy}$).\\
\indent We examine the accretion onto the MBHs by computing the Eddington fraction, $f_{\text{edd}} = \dot{m}/\dot{m}_{\text{edd}}$, for both regions' (i.e. Rarepeak \& Normal) populations. In Figure \ref{fig:edd_frac} we show 2-D histograms of accretion (as a fraction of the Eddington rate) as a function of black hole mass. The accretion rate shown is the maximum rate across the entire redshift range and hence does not truly reflect the mass growth but instead gives a flavour for the maximum instantaneous growth rates. The top panel shows maximum accretion rate from the RP region while the bottom panels show the maximum instantaneous accretion rate from the Normal region.  The left column uses MBHs with initial masses assigned from the P23 relationship, the right hand panel shows initial MBH masses assigned from the RV15 relationship. In orange we show the result when the black holes are positioned consistently at the highest density point while in blue we show the results for when the MBHs follow the dynamics due to the initially nearest DM particle. \\
\indent In all cases the MBH growth increases as a function of MBH mass as expected (since the BHL accretion prescription depend on $\rm{M_{BH}^2}$). Since the P23 relationship gives larger black hole masses as a function of stellar mass we immediately see a greater spread to higher masses in the left hand column. Additionally, allowing the MBHs to consistently sit at the highest density point allows for the most optimistic growth. As a result the most optimistic result is found by looking at the top left panel and focusing on the orange pixels. \\
\indent This can be contrasted with the bottom right panel and focusing on the blue pixels. In this case the spread of masses is much reduced as we employ the RV15 scaling relationship and hence the masses of our seeds are significantly reduced. The bottom right panel also examines the Normal region where the number of host haloes is also significantly suppressed compared to the Rarepeak region. Finally, the blue pixels also use the DM$_{\rm proxy}$ method and therefore account for the dynamics (via a proxy DM particle). This additionally suppresses the accretion. In this (pessimistic) case seeds are effectively light (i.e. $\lesssim 10^3 \ \rm{M_{\odot}}$) and do not grow with average accretion rates at least five orders of magnitude below the Eddington rate. This is the case even when the accretion is boosted, due to unresolved gas, from Eqn.\eqref{eq:boost}

\begin{figure}[htbp]
    \centering
    \includegraphics[width = 0.45\textwidth]{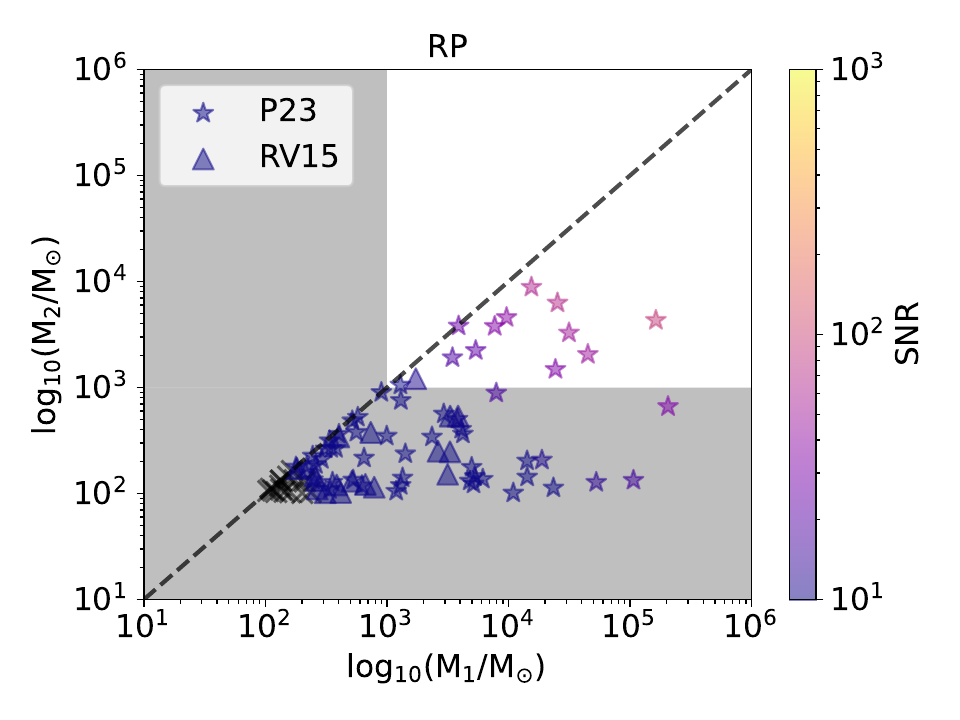}\qquad
    \includegraphics[width = 0.45\textwidth]{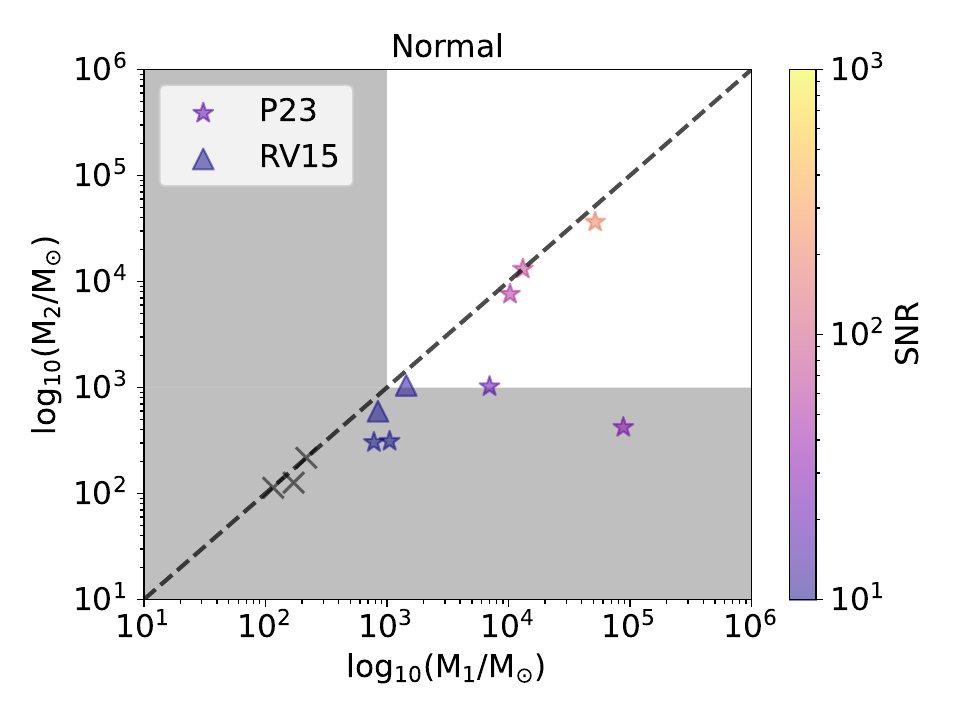}
    \caption{The SNR of the mergers, for the no-delay scenario, found in the Renaissance simulations, with the SNR of the mergers calculated from \cite{Robson_2019}. Due to the lack of growth for the black holes, the $\rho_{max}$ methods and the DM$_{proxy}$ methods are included in the same markers. The black crosses represent the mergers where the SNR could not be calculated and the black dashed line represents the case where the mass ratio between the two merging black holes is 1. The grey shaded regions denote BH masses $<10^3\ \msolar$.}
    \label{fig:mergers}
\end{figure}

\begin{figure}
    \centering
    \includegraphics[width = 0.45\textwidth]{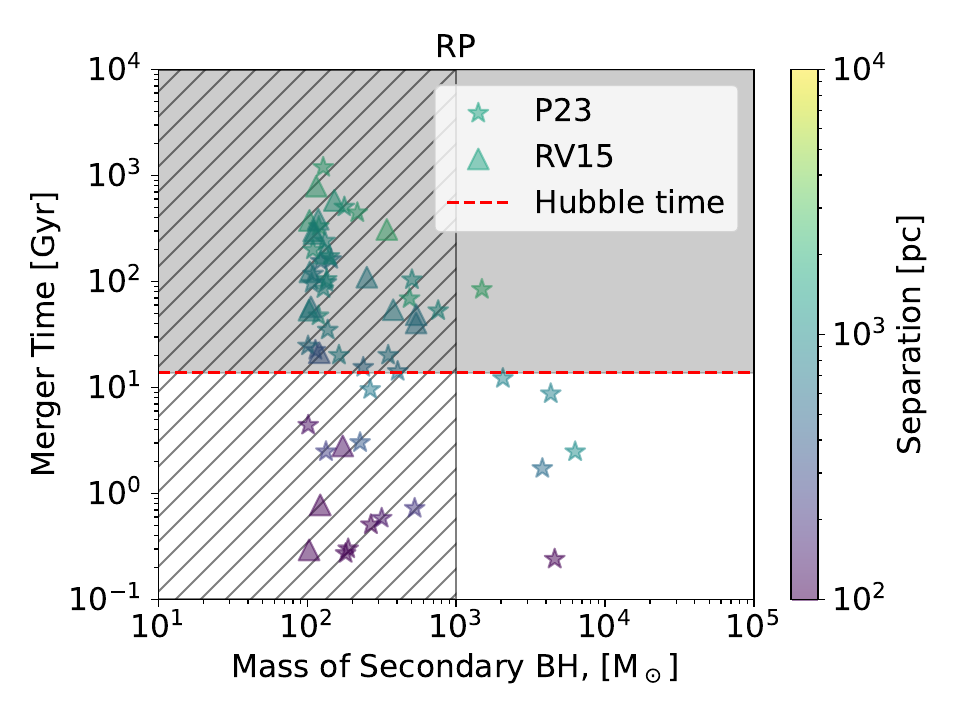}
    \includegraphics[width = 0.45\textwidth]{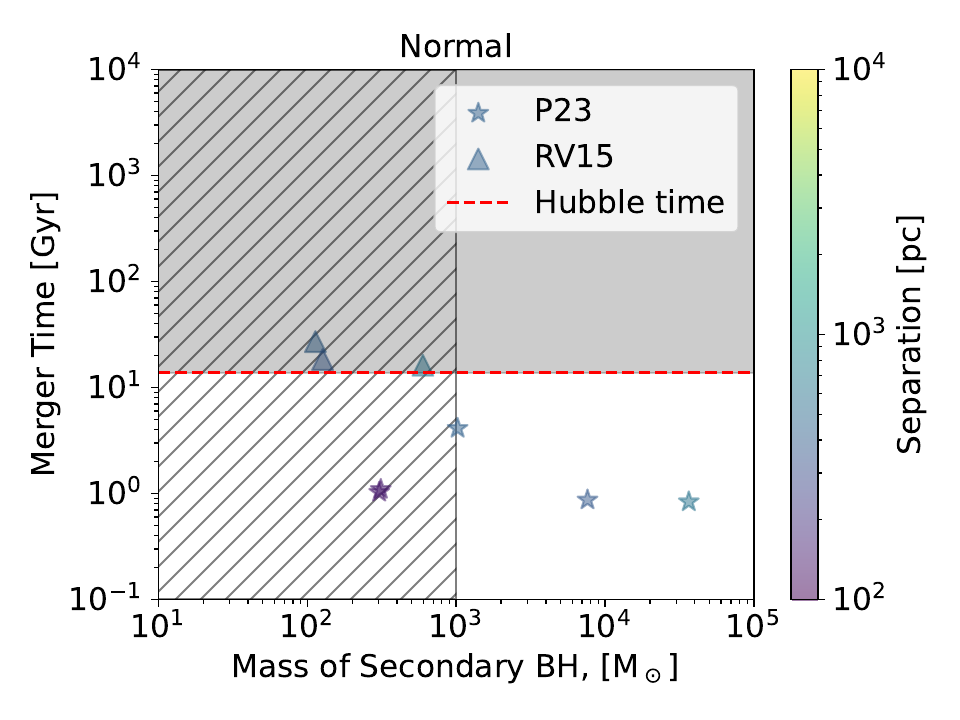}
    \caption{The merger times of the black holes from \renaissance{} with the effects of dynamical friction taken into account. The separation is measured in proper units. As in Figure \ref{fig:mergers}, due to the lack of growth for the MBHs, the $\rho_{max}$ methods and the DM$_{proxy}$ methods are included in the same markers. The grey shaded region denotes mergers that would not be detected due to the fact that they would not have occurred yet. The hatched regions denote BH masses $<10^3\ \msolar$. The region of interest, or at least the region of detectability with LISA, is then the bottom right quadrant. }
    \label{fig:Merger_DF}
\end{figure}

\begin{figure}[tbp]
    \centering
    \includegraphics[width = 0.45\textwidth]{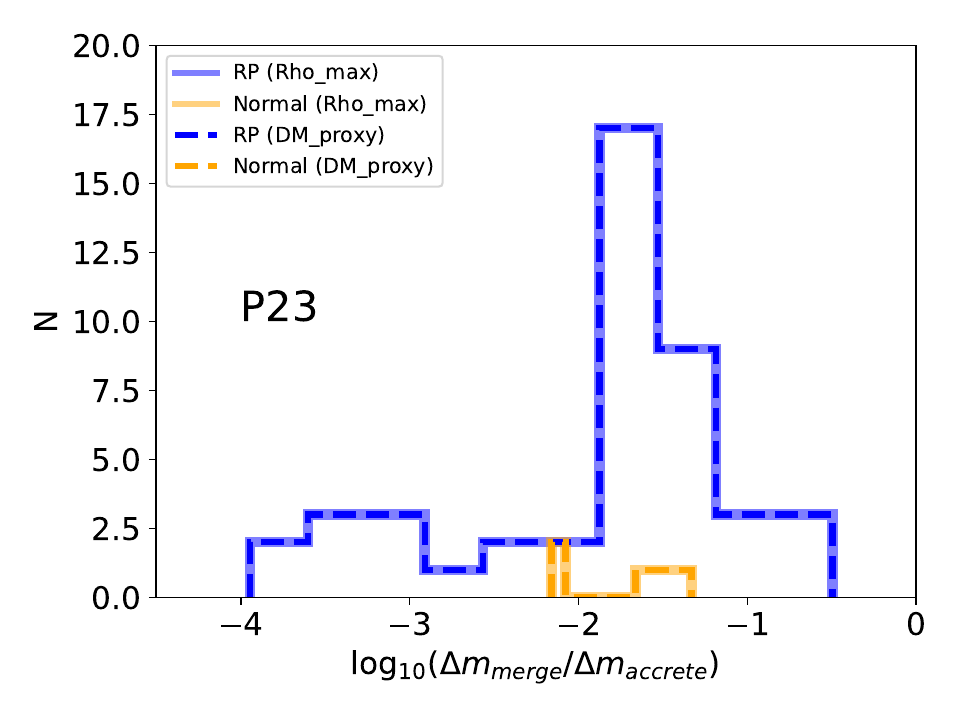}
    \includegraphics[width = 0.45\textwidth]{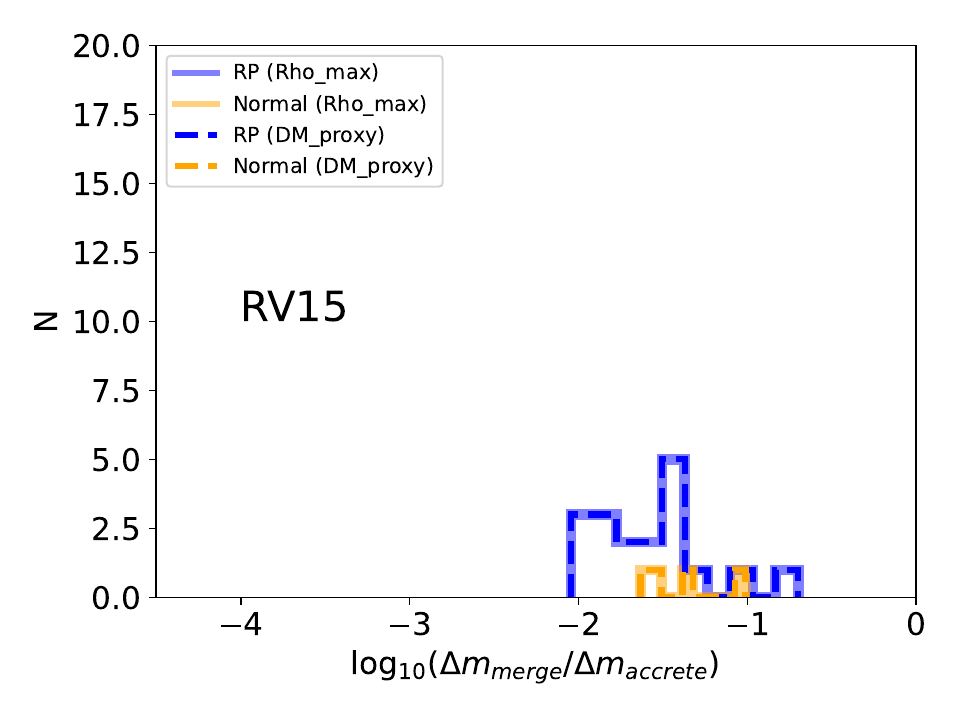}
    \caption{Fraction of MBH growth from mergers compared to growth from accretion for the two regions (RP and Normal). The ratio scale is displayed broadly here to emphasise the point that there is negligible difference between the two regions when it comes to the amount of matter accreted onto the black holes compared to the amount of matter gained from merging. It is noted that the lines for the $\rho_{\text{max}}$ and the $DM_{\text{proxy}}$ method are indistinguishable here}
    \label{fig:mass_growth_fraction}
\end{figure}

\begin{figure*}[htbp]
    \centering
    \includegraphics[width = 0.88\textwidth]{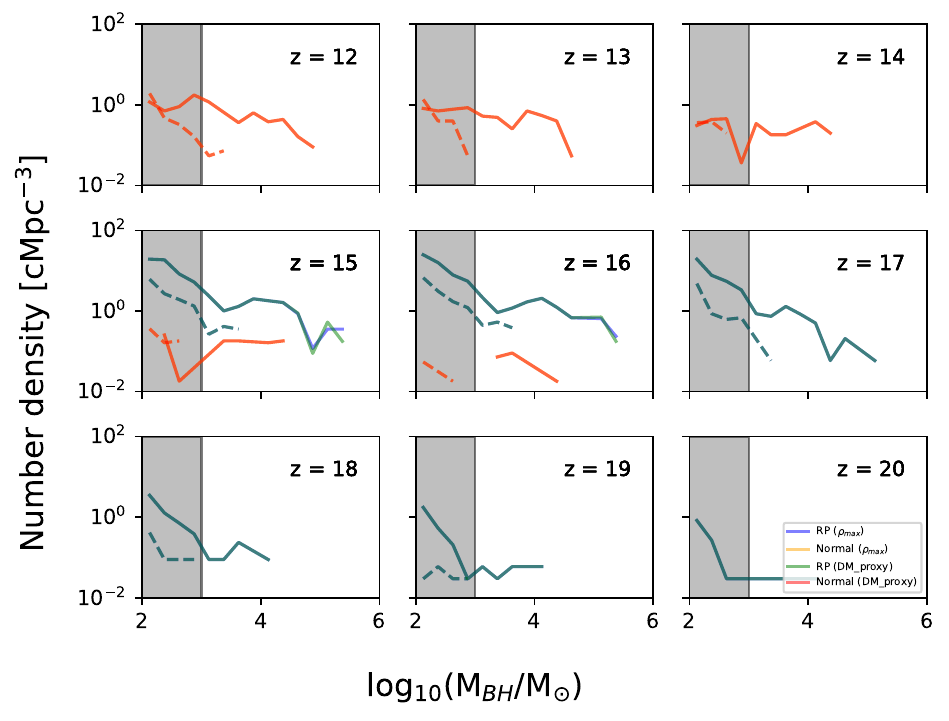}
    \caption{The MBH populations number densities as a function of mass for different redshifts. The solid lines represent the black holes that were seeded with masses derived from P23, and the dashed lines from RV15. The masses here are the result of the growth calculated from the accretion onto the black hole and through merger events with neighbouring halo. It is clear here that a black hole population becomes established earlier in the higher density Rarepeak region compared to the average density region represented by the Normal region. As the population grows the number density follows a power law with the mass dependence. The Rarepeak simulations finish at $z = 15$ which is why there is an absence of black hole population at lower redshift. The grey shaded regions denote BH masses $<10^3\ \msolar$. It is noted that the lines for both accretion methods, in each region, tend to overlap causing a change in the colour of the line compared to the legend.}
    \label{number-density}
\end{figure*}

\subsection{Growth Through Mergers}
In order to examine the growth of MBH via mergers, we investigate two scenarios for a Binary Black Hole (BBH) mergers. In the first case we have the naive situation of the MBH binaries merging immediately after the host halos merge, in the second, more realistic case, we account for delay times to the actual merger. \\
\indent We begin by computing the SNR of signals from BBHs which merge instantaneously. In this case the binaries are merged instantly once the host haloes merge. We compute the SNR expected by LISA using the prescription given in \cite{Robson_2019}. For the SNR calculation we use the masses of the primary and secondary MBHs as well as the redshift that they merge at (without delays). In our model the black holes are also spinless. Adding spin would change our results by a small but not significant amount, \cite{Robson_2019} state that the SNR would only change by at most a factor 2 with spin. In Figure \ref{fig:mergers} we show the SNR for all of the merging black holes computed in our analysis. In the top panel we show the MBH mergers for the Rarepeak region while in the bottom panel we show the results for the Normal region. The Rarepeak being a more clustered region experiences may more halo mergers and hosts more, by number, MBHs. Hence the number of mergers in the Rarepeak regions vastly exceeds those in the Normal region as expected. On top of this the different prescriptions we employ for the MBH seed masses as well as their growth patterns enters. \\
\indent We can see from Figure \ref{fig:mergers} that the low mass black holes, from $10^2 - 10^4\ \msolar$, produce the lowest SNR values as expected and will be extremely challenging to detect with LISA but may be detectable with the Einstein Telescope \citep{Valiante_2021}. Black stars in each panel represent SNR values too small to be reliably calculated. As the MBH masses prior to merger increase, the SNR values increase too. For mergers where both MBH masses are above $10^4$ M$_\odot$, the signal becomes strong enough that it presents the opportunity for these black holes to be candidates for LISA to detect with SNR values reaching upwards of 100. 
However, even in the case where the SNR values are high (SNR > 100) the errors on the luminosity distance will nonetheless present a significant challenge to host galaxy identification and ultimately a strong uncertainty on the masses \citep{kocsisFindingElectromagneticCounterparts2006, sainiPremergerLocalizationIntermediate2022}.\\
\indent Having analysed the SNR from the instantaneous merger case we now move onto determining the merger timescales for a more realistic case i.e. where the MBHs take some additional time to merge following the host halo merger. 
Using the equations described in \S \ref{Sec:DelayTimes} we compute the delay times for the MBH mergers, for MBHs seeded with both scaling relations and both growth methods, and display them in Figure \ref{fig:Merger_DF}. As before the Rarepeak region is displayed in the top panel with the Normal region in the lower panel, for all the growth models we calculated in this analysis. On the x-axis we show the mass of the secondary black hole and on the y-axis the merger time. Each merger 
candidate is coloured by the initial separation. \\
\indent What we see from Figure 
\ref{fig:Merger_DF} is that almost half of the mergers have delay times exceeding the Hubble time particularly for the 
lower mass end of the MBH mass spectrum. There is also a very clear function of separation in the determining the merger timescales. The initial separation between the black holes is calculated by taking the minimum of the distance between the black holes in the snapshot before the host halos are considered merged and the radius of the merged halo.
For the higher MBH masses which are involved in a merger (i.e. those with masses in excess of at least $10^3 \ \msolar$) we see that dynamical friction can drag these MBH to coalescence in less than a Hubble time and with delay times at low as 100 Myr. Our calculations also show a number of light seeds with masses less than $10^3 \ \msolar$ with delay times less than a Hubble time. These lighter seed mergers have short delay times because of the small separations that exist between them just after the galactic merger. It must be noted that the delay time here is based on the parameters and structure of the halo at the time of merger. As the halo evolves through cosmic time it will grow and change in structure, thus affecting the delay time for the mergers. \\
\indent We end by stressing that growth is however primarily driven by accretion rather than mergers. In Figure \ref{fig:mass_growth_fraction} we show the fraction of growth coming from mergers relative to accretion. For both MBHs in the Rarepeak (blue) and Normal regions (yellow) the growth from accretion surpasses the growth from mergers by several orders of magnitude. \\
\indent However, even though mergers account for approximately 1\% of the MBH growth 
they crucially allow us to probe the MBH population demographics and potentially the seeding mechanisms \citep{Sesana_2007, Sesana_2009}. As such it is worth calculating both the 
number of MBH mergers as well as their potential detectability with LISA \citep{LISA_2023, eLISA}. Moreover, the growth via accretion prior to mergers allows for higher SNR values within the LISA band. \\
\indent We also note that as the host halos grow through mergers with smaller halos, there is an increase in the Star Formation Rate in the halo, as shown in \cite{chen_renaissance}. Provided that the halo is massive enough to overcome the effects of stellar feedback, the amount of gas accretion onto the halo also increases. The replenishing of this gas reservoir provides an opportunity for the massive black hole that a halo could host, to gain a boost in the accretion from this gas.\\

\newpage

\subsection{The Number Density of Heavy Seed MBHs}
Finally, having accounted for the initial masses of the heavy seeds, the growth of the black holes via both accretion and mergers we can now calculate the number density of 
heavy seed MBHs as a function of black hole mass and redshift. In Figure \ref{number-density} we show
the number densities of heavy seed MBH. In each panel solid lines represent MBH masses derived from the ``over-massive'' scaling relations from P23 while the dashed lines
are derived using the scaling relations from RV15. The blue and green lines are from the Rarepeak region, while the red and yellow lines from the Normal region. Overall, there is little difference between the number densities of the $\rho_{\text{max}}$ method and the DM$_{\text{proxy}}$ method so there is considerable overlap between the respective lines. The solid lines (being the more optimistic values for the masses) extend to higher masses while the dashed lines (being the more pessimistic values for the masses) extend to less than $\sim 10^4\ \msolar$ in most galaxies simply because our galaxy masses are small. \\
\indent If we consider the $z = 15$ panel (which is the final output from the Rarepeak region and there is also a substantial contribution in the Normal region) we see that even for MBH masses of approximately $10^4\ \msolar$, number densities are as high as $1 \ \rm{cMpc^{-3}}$ for the Rarepeak region and as high as $10^{-1} \rm{cMpc^{-3}}$  for the Normal region. These number densities while admittedly somewhat optimistic as they are derived from the ``over-massive'' relationship show that if heavy seeds can form at high-z in haloes matching conditions similar to what we prescribe then they can easily explain the abundances of AGN currently being detected by JWST \citep{Matthee_2023, Scholtz_2023, Greene_2024}. Moreover, even the most pessimistic case where early MBHs are not over-massive (see top left panel at $z = 12$, dashed line) we still have number densities greater than $10^{-3} \rm{cMpc^{-3}}$ for masses of approximately $10^4 \ \msolar$ (extrapolating the dashed line). Again, if we take this value as a lower limit then it would appear than a subset of haloes with conditions conducive to heavy seed formation can still explain the AGN candidates being put forward by JWST observations - albeit somewhat more marginally \citep{Greene_2024}. \\
\indent Taking our predictions at face value our calculations predict number densities for MBHs with masses of approximately $10^4\ \msolar$ from $10^{-2} \ \rm{cMpc^{-3}}$ up to $1 \ \rm{cMpc^{-3}}$ at $z = 15$. Our ability to predict number densities at lower redshifts is hampered by our underlying simulation suite which does not provide statistics for us at significantly lower redshifts. However, if we take the outputs at $z = 12$ from the Normal region and again assume an over-massive population then the number densities hold and we predict
number densities for MBHs with masses of approximately $10^4\ \msolar$ again up to $1 \ \rm{cMpc^{-3}}$ at z = 12. The number densities for more massive objects (as is being detected as high-z AGN by JWST) will be lower by a couple of orders of magnitude but this nonetheless fits well with current AGN candidate estimates \citep{Matthee_2023, Scholtz_2023, Greene_2024}.

\subsection{Comparing our Number Density Rates to the Existing Literature}
\noindent Calculating the number density of MBHs in the high-z Universe is a hugely important goal in modern astrophysics. The number density itself is a key parameter in further calculating the active fraction of MBHs in galaxies \citep[e.g.][]{Pacucci_2021} and in calculations of merger rates of MBHs potentially detectable by the next generation of space based gravitational wave observatories \citep{LISA_2023}. \\
\indent Over the last two decades many studies have attempted to calculate the overall number density values as a function of redshift as we have done here. The initial calculations from the early part of the last decade came from \cite{Dijkstra_2014} and \cite{Agarwal_2012}. Their calculation were primarily driven by assuming that MBH form in haloes subject to a supercritical Lyman-Werner(LW) flux. Depending on the strength of the flux assumed the values calculated by both authors are between approximately $10^{-10}$ to $10^{-6}$ cMpc$^{-3}$ at z $\sim 10$ and so many orders of magnitude more pessimistic than the values found here (see Figure \ref{number-density}). \\
\indent Higher values for the expected number density of MBHs can be obtained by reducing the value of the super-critical flux required. The values quoted previously are for J$_{\rm{crit}} = 300 \ \rm{J_{21}}$\footnote{where $\rm{J_{21}}$ is in units of $10^{-21} \rm{erg} \ cm^{-2} \ Hz^{-1} \ s^{-1} \ str^{-1}$} but by lowering the threshold to J$_{\rm{crit}} = 30 \ \rm{J_{21}}$ 
values aligning closer to this study's can be obtained. However, the majority of research in the literature suggests that the higher threshold of J$_{\rm{crit}} = 300 \ \rm{J_{21}}$ (or higher is required) \citep{Shang_2010, Visbal_2014a,Regan_2014, Latif_2015,Regan_2016, Regan_2017}. \\\indent \cite{Habouzit_2016} similarly found that number densities of order  $10^{-6}$ cMpc$^{-3}$ at z $\sim 10$ result from a threshold J$_{\rm{crit}}$ = 300 when analysing output from several cosmological hydrodynamical simulations that cover a large range of box sizes and resolutions. Lessening the metallicity constraint to incorporate metal-poor rather than metal-free haloes improves the number densities somewhat \citep{Dunn_2018, Chon_2020, Regan_2020a} bringing number density values up to approximately $10^{-4}$ cMpc$^{-3}$ at z $\sim 10$ \citep{Visbal_2014,Lupi_2021, Trinca_2022, Spinoso_2024}. \\
\indent However, such models are always somewhat sensitive to the parameters used to analyse the data as no model can yet follow, in a cosmological context, the formation of MBH seeds self-consistently. Some models therefore predict somewhat higher number densities even for the LW pathway \citep[e.g.][]{Chiaki_2023} but the consensus of the bulk of the literature appears to suggest number densities of approximately $10^{-6}$ cMpc$^{-3}$ to $10^{-4}$ cMpc$^{-3}$at $z \sim 10$ for reasonable values of J$_{\rm{crit}}$ (i.e. J$_{\rm{crit}} \gtrsim 300 \ \rm{J}_{21}$).\\
\indent The values we find, through appealing to the so-called `rapid assembly model' \citep[e.g.][]{Yoshida_2003, Wise_2019}, are at least two to three orders of magnitude higher than the more optimistic LW channel results (and in fact are closer to some semi-analytic models based on very efficient growth of light seeds \citep[e.g.][]{Trinca_2022,Spinoso_2024}). As discussed in \S XYZ our values are founded on the rapid growth of the host halo predominantly following on from the results of \cite{Wise_2019}. A comparable study by \cite{Lupi_2021} 
used a semi-analytical treatment on top of n-body only calculations of a quasar progenitor environment to calculate number densities in a similar way. \cite{Lupi_2021} found number density values of approximately 0.1 cMpc$^{-3}$ at $z \sim 10$ (after accounting for the impact of their overdense realisation). This value compares extremely well to our estimate (see again Figure \ref{number-density}). \\
\indent In general terms a heavy seed model relying solely on the influence of a LW background will struggle to meet the number density demands currently being set by JWST \citep[e.g.][]{Matthee_2023, Greene_2024}. Our methodology, based primarily on rapid halo growth, achieves rates that match and exceed those required by current observations with number densities reaching well above $10^{-2}$ cMpc$^{-3}$ at $z \sim 10$ (for MBHs with masses of approximately $10^4$ M$_{\odot}$).

\subsection{LISA Event Rates from Mergers of Heavy Seed Massive Black Holes}
\begin{figure}
    \centering
    \includegraphics[width = 0.45\textwidth]{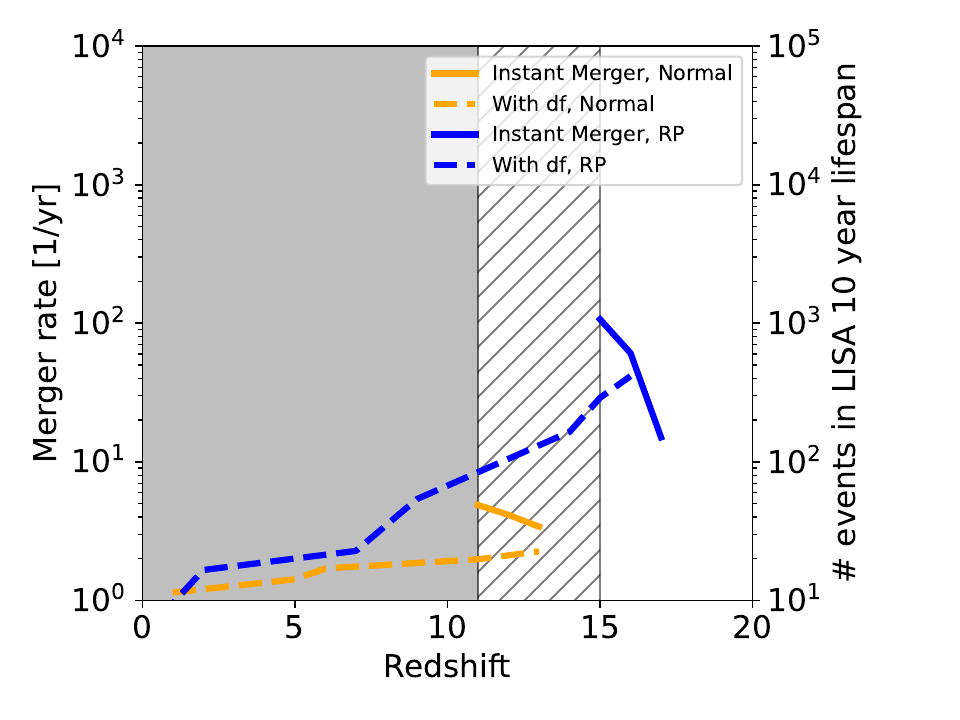}
    \caption{The merger rate of the massive black holes from \renaissance{}, for both the instant mergers and the mergers with time delay due to dynamical friction. The hatched dashed region denotes the redshift regime where the Rarepeak is unable to simulate, whilst the solid grey region denotes where the RP and Normal region are unable to simulate. In the lifetime of LISA (~10 yrs) we should expect to see $\sim 80$ mergers from MBH binaries using the results from the Normal region.}
    \label{fig:merger-rate}
\end{figure}

\noindent Using the mergers that are predicted at high-redshift with Renaissance, we are able to make an estimation for the merger rate of these MBHs that will then be observed during LISA's 4--10 year lifespan. We calculate the merger rate using the following formula:
\begin{equation}
    \frac{\dd N}{\dd z \dd t_{\text{obs}}} = 4\pi c r_c^2 \frac{\dd N}{\dd z \dd V}
\end{equation}
where $r_c$ is the comoving distance to the merger events and $\dd N/\dd z \dd V$ is the number density of mergers per redshift, per comoving volume \citep{haehneltLowfrequencyGravitationalWaves1994, ciardiExpectedNumberFlux2000, chakrabortyProbingSixMassive2023}. We compute the merger rate for both the instant mergers (after the merging of the host galaxies) and the mergers that are time delayed due to the effects of dynamical friction. We show the merger rate calculation in Figure \ref{fig:merger-rate}.
\par It must be noted that the merger rate appears to decrease for the time delayed case because the Normal region does not simulate past $z = 11.6$, and therefore the mergers of any MBHs after this time is not simulated. Using Figure \ref{fig:merger-rate}, we can see that in the proposed 10 year lifetime of LISA should expect to see around $\sim 80$ mergers, using the predictions form the Normal region. With the RP predictions we expect to see around $10^2 - 10^3$ mergers in LISA's lifetime. With the addition of dynamical friction into the merger model we can expect to see signals from binaries formed in the early Universe, merging at redshifts closer to the local Universe.
\subsection{Detection of High-z Black Holes and Comparison with Other Semi-analytical Models}

The high redshift nature of the black holes modelled in this analysis makes them difficult to detect using current observation techniques. While JWST is continuing to make measurements of high-z AGN, the extreme nature of these active black holes are potentially causing bias in the predicted mass spectrum of MBHs in the early Universe \citep{JunyaoTipOfTheIceberg}. The low accretion values of our black holes will lead to low bolometric luminosity and thus the need for other detection techniques becomes greater. With the future gravitational wave detector LISA due to be launched in the 2030's, it will present us with the opportunity to detect gravitational wave signatures from merging black holes like the ones modelled here. The ESA led space-based gravitational wave detector is designed to detect mergers from black holes in the intermediate mass range, up to and at very high redshifts ($z > 15$). LISA will not be alone in the search for higher mass black hole binaries, the Chinese led missions Taiji \citep{Taiji} and Tianqin \citep{Tianqin} will operate with the same principles as LISA, while the ground based Einstein telescope \citep{EinsteinTelescope} and Cosmic Explorer \citep{Cosmic_explorer} will join the search for black hole mergers up to the cosmic dark ages.

Other semi-analytical and simulation models have explored the extent to which GW observatories will explore the the high-z population of massive black holes \citep[e.g.][]{sesanaImprintMassiveBlack2007, tanakaASSEMBLYSUPERMASSIVEBLACK2009,
dayalHierarchicalAssemblyGalaxies2019,valianteUnveilingEarlyBlack2020}. The general consensus is that if we are able to constrain the parameters from black hole mergers and analyse a merger rate (whose predicted values range from $10^1 - 10^3 \rm{ yr}^{-1}$ with LISA) using the synergy of various gravitational wave detectors, we should be able to constrain the massive black hole formation models which dominate the early Universe.



\section{Summary \& Conclusions}
\label{sec:conclusions}


\par 
\noindent In this paper, we set out to study the halo environments potentially conducive to heavy seeds formation in the early Universe via post process modelling in the \renaissance{} simulations. In order for a halo to host a heavy seed, it would need to satisfy four criteria: (i) reaching a mass needed for atomic cooling, (ii) a high inflow rate ($\dot{M} > 0.1$ \msolaryr), (iii) low metallicity ($Z < 10^{-3} \zsolar$) and (iv) compactness. In effect we find that the high mass inflow rate is the strictest criteria with the other criteria more easily attained. Because the final redshifts of each region are quite high, they don't fully capture the end of massive black hole formation, as can be seen in fig.\ref{fig:conditions}, therefore we are underestimating the mergers that will happen after the end times of \renaissance{}. Nonetheless it is important to include these mergers in our black hole formation models as they will contribute to the signal that LISA will detect.\\
\indent Once the host haloes are identified, we assign masses to the black holes in one of two ways. On the one hand we assign MBH masses from a local relationship between galaxy masses and black holes masses from RV15 and on the other hand we utilise a more recent relationship specific to high-z MBH obtained from JWST data (P23). The latter relationship predicts that high-z black holes are ``over-massive'' with respect to their host galaxy. Following seeding the black holes were then allowed to grow, both via accretion and mergers with each other. We based the accretion formalism on the Bondi-Hoyle-Lyttleton model and calculate the actual accretion rate using parameters obtained from the underlying \renaissance{} simulations. For accretion we investigate two models, one where the MBHs are effectively pinned to the highest density point, $\rho_{max}$, and secondly where the MBH moves within the gravitational potential of the host galaxy. Because the accretion methods are applied to snapshots in \renaissance, they are instantaneous values and may not accurately track the growth of black holes over the simulation time. Therefore, we may be over or under-predicting the final simulated masses of the MBH population. However, the predicted observed luminosity will not be affected by these calculations. \\
\indent Finally, MBH mergers are also included in our analysis both from the point of view of growth but also from the point of view of observability with future gravitational wave observatories.  For analysing MBH mergers we again consider two scenarios. In the naive case we assume that the BBHs merge instantly as part of the host halo merger and secondly we consider the case where the black holes must coalesce via dynamical friction. The latter case can, in some instances, result in significant delay times between the host galaxy merger and the BBH merger. Our main conclusions are:
\begin{enumerate}
 \item Using a heavy seed model for populating dark matter haloes we find that the number density of MBHs from this channel can be as high as $1 \ \rm{cMpc^{-3}}$ at $z \sim 10 - 15$ in the most optimistic case. In a more pessimistic scenario this value can drop by three orders of magnitude, for masses in the range of $10^2 - 10^4 \msolar$. However, even at this lower level of $10^{-2} \ \rm{cMpc^{-3}}$ at $z \sim 10 - 15$
 a heavy seed model can still explain the number of AGN candidates currently being detected by JWST. Moreover, what this tells us is that host galaxy environments with high inflow rates, low metallicity and high compactness exist in sufficient numbers to explain the JWST candidate AGN. 
\item MBH growth following seeding is in general sub-Eddington. Only the most massive MBHs (M$_{\rm MBH} \gtrsim 10^{4-5}\ \msolar$) can grow at values approaching the Eddington rate. This also only occurs for those MBHs which are pinned to the highest density regions. MBHs which undergo (random) motions around the galactic nucleus show strongly suppressed growth. 
\item MBH growth comes primarily, by several orders of magnitude, from gas accretion rather than MBH mergers.
\item MBH mergers are in principle observable with LISA for primary black hole masses in excess of $10^4\ \msolar$ and secondary black hole masses in excess of $10^3\ \msolar$. In these cases SNR values in excess of 100 are viable with delay times (assuming they are shorter than the age of the Universe) increasing the SNR values and potentially also easing difficulties with host galaxy detection. 
\item We calculate LISA event rates of both the Normal and Rarepeak regions and find that the high redshift regime will provide numerous candidates for LISA detection in its proposed 10 year lifespan, with $\sim 80$ mergers expected to be observed using the merger predictions from the Normal region (with $\sim 10^3$ mergers predicted from the RP region). If LISA is able to detect the signals from these mergers, we will be able to further constrain the formation processes of the black hole populations at this epoch. 
\end{enumerate}

\section*{Acknowledgements}
\noindent JM acknowledges the support from the John \& Pat Hume Doctoral Awards Scholarship (Hume 2021-22). JR acknowledges support from the Royal Society and Science Foundation Ireland under grant number 
 URF\textbackslash R1\textbackslash 191132. JR acknowledges support from the Irish Research Council Laureate programme under grant number IRCLA/2022/1165. BS acknowledges support by STFC Consolidated Grant RA5496. JW acknowledges support by NSF grant AST-2108020 and NASA grants 80NSSC20K0520 and 80NSSC21K1053. BWO acknowledges support from NSF grants AAG-1908109 and AAG-2106575, NASA ATP grants NNX15AP39G and 80NSSC18K1105, and NASA TCAN grant 80NSSC21K1053.

\bibliographystyle{mn2e}
\bibliography{references, MyLibrary}

\begin{thebibliography}{159}
\expandafter\ifx\csname natexlab\endcsname\relax\def\natexlab#1{#1}\fi

\bibitem[{{Agarwal} {et~al}\mbox{.}(2012){Agarwal}, {Khochfar}, {Johnson}, {Neistein}, {Dalla Vecchia}, \& {Livio}}]{Agarwal_2012}
{Agarwal} B., {Khochfar} S., {Johnson} J.~L., {Neistein} E., {Dalla Vecchia} C., {Livio} M., 2012, \mnras, 425, 2854

\bibitem[{{Alexander} \& {Natarajan}(2014)}]{Alexander_2014}
{Alexander} T., {Natarajan} P., 2014, Science, 345, 1330

\bibitem[{{Alvarez}, {Wise} \& {Abel}(2009){Alvarez}, {Wise}, \& {Abel}}]{Alvarez_2009}
{Alvarez} M.~A., {Wise} J.~H., {Abel} T., 2009, \apjl, 701, L133

\bibitem[{{Amaro-Seoane} {et~al}\mbox{.}(2023){Amaro-Seoane}, {Andrews}, {Arca Sedda}, {Askar}, {Baghi}, {Balasov}, {Bartos}, {Bavera}, {Bellovary}, {Berry}, {Berti}, {Bianchi}, {Blecha}, {Blondin}, {Bogdanovi{\'c}}, {Boissier}, {Bonetti}, {Bonoli}, {Bortolas}, {Breivik}, {Capelo}, {Caramete}, {Cattorini}, {Charisi}, {Chaty}, {Chen}, {Chru{\'s}li{\'n}ska}, {Chua}, {Church}, {Colpi}, {D'Orazio}, {Danielski}, {Davies}, {Dayal}, {De Rosa}, {Derdzinski}, {Destounis}, {Dotti}, {Dutan}, {Dvorkin}, {Fabj}, {Foglizzo}, {Ford}, {Fouvry}, {Franchini}, {Fragos}, {Fryer}, {Gaspari}, {Gerosa}, {Graziani}, {Groot}, {Habouzit}, {Haggard}, {Haiman}, {Han}, {Istrate}, {Johansson}, {Khan}, {Kimpson}, {Kokkotas}, {Kong}, {Korol}, {Kremer}, {Kupfer}, {Lamberts}, {Larson}, {Lau}, {Liu}, {Lloyd-Ronning}, {Lodato}, {Lupi}, {Ma}, {Maccarone}, {Mandel}, {Mangiagli}, {Mapelli}, {Mathis}, {Mayer}, {McGee}, {McKernan}, {Miller}, {Mota}, {Mumpower}, {Nasim}, {Nelemans}, {Noble}, {Pacucci}, {Panessa}, {Paschalidis}, {Pfister}, {Porquet},
  {Quenby}, {Ricarte}, {R{\"o}pke}, {Regan}, {Rosswog}, {Ruiter}, {Ruiz}, {Runnoe}, {Schneider}, {Schnittman}, {Secunda}, {Sesana}, {Seto}, {Shao}, {Shapiro}, {Sopuerta}, {Stone}, {Suvorov}, {Tamanini}, {Tamfal}, {Tauris}, {Temmink}, {Tomsick}, {Toonen}, {Torres-Orjuela}, {Toscani}, {Tsokaros}, {Unal}, {V{\'a}zquez-Aceves}, {Valiante}, {van Putten}, {van Roestel}, {Vignali}, {Volonteri}, {Wu}, {Younsi}, {Yu}, {Zane}, {Zwick}, {Antonini}, {Baibhav}, {Barausse}, {Bonilla Rivera}, {Branchesi}, {Branduardi-Raymont}, {Burdge}, {Chakraborty}, {Cuadra}, {Dage}, {Davis}, {de Mink}, {Decarli}, {Doneva}, {Escoffier}, {Gandhi}, {Haardt}, {Lousto}, {Nissanke}, {Nordhaus}, {O'Shaughnessy}, {Portegies Zwart}, {Pound}, {Schussler}, {Sergijenko}, {Spallicci}, {Vernieri}, \& {Vigna-G{\'o}mez}}]{LISA_2023}
{Amaro-Seoane} P. {et~al.}, 2023, Living Reviews in Relativity, 26, 2

\bibitem[{{Antonini}, {Gieles} \& {Gualandris}(2019){Antonini}, {Gieles}, \& {Gualandris}}]{Antonini_2019}
{Antonini} F., {Gieles} M., {Gualandris} A., 2019, \mnras, 486, 5008

\bibitem[{{Arca Sedda}, {Amaro Seoane} \& {Chen}(2021){Arca Sedda}, {Amaro Seoane}, \& {Chen}}]{Arca-Sedda_2021}
{Arca Sedda} M., {Amaro Seoane} P., {Chen} X., 2021, \aap, 652, A54

\bibitem[{Begelman, Volonteri \& Rees(2006)Begelman, Volonteri, \& Rees}]{begelmanFormationSupermassiveBlack2006}
Begelman M.~C., Volonteri M., Rees M.~J., 2006, Monthly Notices of the Royal Astronomical Society, 370, 289

\bibitem[{Behroozi, Wechsler \& Wu(2012)Behroozi, Wechsler, \& Wu}]{Behroozi_2012}
Behroozi P.~S., Wechsler R.~H., Wu H.-Y., 2012, The Astrophysical Journal, 762, 109

\bibitem[{Behroozi {et~al}\mbox{.}(2012)Behroozi, Wechsler, Wu, Busha, Klypin, \& Primack}]{Behroozi_Wechsler_Wu_Busha_Klypin_Primack_2012}
Behroozi P.~S., Wechsler R.~H., Wu H.-Y., Busha M.~T., Klypin A.~A., Primack J.~R., 2012, The Astrophysical Journal, 763, 18

\bibitem[{Bennett {et~al}\mbox{.}(2023)Bennett, Sijacki, Costa, Laporte, \& Witten}]{bennettGrowthGargantuanBlack2023}
Bennett J.~S., Sijacki D., Costa T., Laporte N., Witten C., 2023, Monthly Notices of the Royal Astronomical Society, 527, 1033

\bibitem[{Binney \& Tremaine(2008)}]{binneyGalacticDynamicsSecond2008}
Binney J., Tremaine S., 2008, Galactic {Dynamics}: {Second} {Edition}. Publication Title: Galactic Dynamics: Second Edition ADS Bibcode: 2008gady.book.....B

\bibitem[{Bleuler \& Teyssier(2014)}]{volume_integral}
Bleuler A., Teyssier R., 2014, Monthly Notices of the Royal Astronomical Society, 445, 4015

\bibitem[{Boekholt {et~al}\mbox{.}(2018)Boekholt, Schleicher, Fellhauer, Klessen, Reinoso, Stutz, \& Haemmerlé}]{Boekholt_Schleicher_Fellhauer_Klessen_Reinoso_Stutz_Haemmerlé_2018}
Boekholt T. C.~N., Schleicher D. R.~G., Fellhauer M., Klessen R.~S., Reinoso B., Stutz A.~M., Haemmerlé L., 2018, Monthly Notices of the Royal Astronomical Society, 476, 366–380

\bibitem[{{Bogd{\'a}n} {et~al}\mbox{.}(2023){Bogd{\'a}n}, {Goulding}, {Natarajan}, {Kov{\'a}cs}, {Tremblay}, {Chadayammuri}, {Volonteri}, {Kraft}, {Forman}, {Jones}, {Churazov}, \& {Zhuravleva}}]{Bogdan_2023}
{Bogd{\'a}n} {\'A}. {et~al.}, 2023, Nature Astronomy

\bibitem[{{Bondi}(1952)}]{Bondi_1952}
{Bondi} H., 1952, \mnras, 112, 195

\bibitem[{Booth \& Schaye(2009)}]{boothCosmologicalSimulationsGrowth2009}
Booth C.~M., Schaye J., 2009, Monthly Notices of the Royal Astronomical Society, 398, 53

\bibitem[{Bromm {et~al}\mbox{.}(2001)Bromm, Ferrara, Coppi, \& Larson}]{brommFragmentationPreenrichedPrimordial2001}
Bromm V., Ferrara A., Coppi P., Larson R., 2001, Monthly Notices of the Royal Astronomical Society, 328, 969

\bibitem[{{Brummel-Smith} {et~al}\mbox{.}(2019){Brummel-Smith}, {Bryan}, {Butsky}, {Corlies}, {Emerick}, {Forbes}, {Fujimoto}, {Goldbaum}, {Grete}, {Hummels}, {Kim}, {Koh}, {Li}, {Li}, {Li}, {OShea}, {Peeples}, {Regan}, {Salem}, {Schmidt}, {Simpson}, {Smith}, {Tumlinson}, {Turk}, {Wise}, {Abel}, {Bordner}, {Cen}, {Collins}, {Crosby}, {Edelmann}, {Hahn}, {Harkness}, {Harper-Clark}, {Kong}, {Kritsuk}, {Kuhlen}, {Larrue}, {Lee}, {Meece}, {Norman}, {Oishi}, {Paschos}, {Peruta}, {Razoumov}, {Reynolds}, {Silvia}, {Skillman}, {Skory}, {So}, {Tasker}, {Wagner}, {Wang}, {Xu}, \& {Zhao}}]{Enzo_2019}
{Brummel-Smith} C. {et~al.}, 2019, The Journal of Open Source Software, 4, 1636

\bibitem[{{Bryan} {et~al}\mbox{.}(2014){Bryan}, {Norman}, {O'Shea}, {Abel}, {Wise}, {Turk}, \& {The Enzo Collaboration}}]{Enzo_2014}
{Bryan} G.~L., {Norman} M.~L., {O'Shea} B.~W., {Abel} T., {Wise} J.~H., {Turk} M.~J., {The Enzo Collaboration}, 2014, \apjs, 211, 19

\bibitem[{Chakraborty {et~al}\mbox{.}(2023)Chakraborty, Gallerani, Zana, Sesana, Valentini, Izquierdo-Villalba, Di~Mascia, Vito, \& Barai}]{chakrabortyProbingSixMassive2023}
Chakraborty S. {et~al.}, 2023, Monthly Notices of the Royal Astronomical Society, 523, 758, aDS Bibcode: 2023MNRAS.523..758C

\bibitem[{Chandrasekhar(1943)}]{chandrasekharDynamicalFrictionGeneral1943}
Chandrasekhar S., 1943, The Astrophysical Journal, 97, 255

\bibitem[{Chattopadhyay {et~al}\mbox{.}(2023)Chattopadhyay, Stegmann, Antonini, Barber, \& Romero-Shaw}]{Chattopadhyay_Stegmann_Antonini_Barber_Romero-Shaw_2023}
Chattopadhyay D., Stegmann J., Antonini F., Barber J., Romero-Shaw I.~M., 2023, Monthly Notices of the Royal Astronomical Society, 526, 4908–4928

\bibitem[{{Chen} {et~al}\mbox{.}(2014){Chen}, {Wise}, {Norman}, {Xu}, \& {O'Shea}}]{chen_renaissance}
{Chen} P., {Wise} J.~H., {Norman} M.~L., {Xu} H., {O'Shea} B.~W., 2014, \apj, 795, 144

\bibitem[{{Chiaki} {et~al}\mbox{.}(2023){Chiaki}, {Chon}, {Omukai}, {Trinca}, {Schneider}, \& {Valiante}}]{Chiaki_2023}
{Chiaki} G., {Chon} S., {Omukai} K., {Trinca} A., {Schneider} R., {Valiante} R., 2023, \mnras, 521, 2845

\bibitem[{Chon \& Omukai(2020)}]{chonSupermassiveStarFormation2020}
Chon S., Omukai K., 2020, Monthly Notices of the Royal Astronomical Society, 494, 2851

\bibitem[{{Chon} \& {Omukai}(2020)}]{Chon_2020}
{Chon} S., {Omukai} K., 2020, \mnras, 494, 2851

\bibitem[{Ciardi \& Loeb(2000)}]{ciardiExpectedNumberFlux2000}
Ciardi B., Loeb A., 2000, The Astrophysical Journal, 540, 687

\bibitem[{Das {et~al}\mbox{.}(2021)Das, Schleicher, Basu, \& Boekholt}]{Das_Schleicher_Basu_Boekholt_2021}
Das A., Schleicher D. R.~G., Basu S., Boekholt T. C.~N., 2021, Monthly Notices of the Royal Astronomical Society, 505, 2186–2194

\bibitem[{{Dav{\'e}} {et~al}\mbox{.}(2019){Dav{\'e}}, {Angl{\'e}s-Alc{\'a}zar}, {Narayanan}, {Li}, {Rafieferantsoa}, \& {Appleby}}]{SIMBA}
{Dav{\'e}} R., {Angl{\'e}s-Alc{\'a}zar} D., {Narayanan} D., {Li} Q., {Rafieferantsoa} M.~H., {Appleby} S., 2019, \mnras, 486, 2827

\bibitem[{Dayal {et~al}\mbox{.}(2019)Dayal, Rossi, Shiralilou, Piana, Choudhury, \& Volonteri}]{dayalHierarchicalAssemblyGalaxies2019}
Dayal P., Rossi E.~M., Shiralilou B., Piana O., Choudhury T.~R., Volonteri M., 2019, Monthly Notices of the Royal Astronomical Society, 486, 2336

\bibitem[{{Devecchi} \& {Volonteri}(2009)}]{Devecchi_2008}
{Devecchi} B., {Volonteri} M., 2009, \apj, 694, 302

\bibitem[{{Dijkstra}, {Ferrara} \& {Mesinger}(2014){Dijkstra}, {Ferrara}, \& {Mesinger}}]{Dijkstra_2014}
{Dijkstra} M., {Ferrara} A., {Mesinger} A., 2014, \mnras, 442, 2036

\bibitem[{Dubois {et~al}\mbox{.}(2015)Dubois, Volonteri, Silk, Devriendt, Slyz, \& Teyssier}]{duboisBlackHoleEvolution2015}
Dubois Y., Volonteri M., Silk J., Devriendt J., Slyz A., Teyssier R., 2015, Monthly Notices of the Royal Astronomical Society, 452, 1502

\bibitem[{{Dunn} {et~al}\mbox{.}(2018){Dunn}, {Bellovary}, {Holley-Bockelmann}, {Christensen}, \& {Quinn}}]{Dunn_2018}
{Dunn} G., {Bellovary} J., {Holley-Bockelmann} K., {Christensen} C., {Quinn} T., 2018, \apj, 861, 39

\bibitem[{{eLISA Consortium} {et~al}\mbox{.}(2013){eLISA Consortium}, {Amaro Seoane}, {Aoudia}, {Audley}, {Auger}, {Babak}, {Baker}, {Barausse}, {Barke}, {Bassan}, {Beckmann}, {Benacquista}, {Bender}, {Berti}, {Bin{\'e}truy}, {Bogenstahl}, {Bonvin}, {Bortoluzzi}, {Brause}, {Brossard}, {Buchman}, {Bykov}, {Camp}, {Caprini}, {Cavalleri}, {Cerdonio}, {Ciani}, {Colpi}, {Congedo}, {Conklin}, {Cornish}, {Danzmann}, {de Vine}, {DeBra}, {Dewi Freitag}, {Di Fiore}, {Diaz Aguilo}, {Diepholz}, {Dolesi}, {Dotti}, {Fern{\'a}ndez Barranco}, {Ferraioli}, {Ferroni}, {Finetti}, {Fitzsimons}, {Gair}, {Galeazzi}, {Garcia}, {Gerberding}, {Gesa}, {Giardini}, {Gibert}, {Grimani}, {Groot}, {Guzman Cervantes}, {Haiman}, {Halloin}, {Heinzel}, {Hewitson}, {Hogan}, {Holz}, {Hornstrup}, {Hoyland}, {Hoyle}, {Hueller}, {Hughes}, {Jetzer}, {Kalogera}, {Karnesis}, {Kilic}, {Killow}, {Klipstein}, {Kochkina}, {Korsakova}, {Krolak}, {Larson}, {Lieser}, {Littenberg}, {Livas}, {Lloro}, {Mance}, {Madau}, {Maghami}, {Mahrdt}, {Marsh}, {Mateos},
  {Mayer}, {McClelland}, {McKenzie}, {McWilliams}, {Merkowitz}, {Miller}, {Mitryk}, {Moerschell}, {Mohanty}, {Monsky}, {Mueller}, {M{\"u}ller}, {Nelemans}, {Nicolodi}, {Nissanke}, {Nofrarias}, {Numata}, {Ohme}, {Otto}, {Perreur-Lloyd}, {Petiteau}, {Phinney}, {Plagnol}, {Pollack}, {Porter}, {Prat}, {Preston}, {Prince}, {Reiche}, {Richstone}, {Robertson}, {Rossi}, {Rosswog}, {Rubbo}, {Ruiter}, {Sanjuan}, {Sathyaprakash}, {Schlamminger}, {Schutz}, {Sch{\"u}tze}, {Sesana}, {Shaddock}, {Shah}, {Sheard}, {Sopuerta}, {Spector}, {Spero}, {Stanga}, {Stebbins}, {Stede}, {Steier}, {Sumner}, {Sun}, {Sutton}, {Tanaka}, {Tanner}, {Thorpe}, {Tr{\"o}bs}, {Tinto}, {Tu}, {Vallisneri}, {Vetrugno}, {Vitale}, {Volonteri}, {Wand}, {Wang}, {Wanner}, {Ward}, {Ware}, {Wass}, {Weber}, {Yu}, {Yunes}, \& {Zweifel}}]{eLISA}
{eLISA Consortium} {et~al.}, 2013, ArXiv e-prints 1305.5720

\bibitem[{Escala(2021)}]{Escala_2021}
Escala A., 2021, The Astrophysical Journal, 908, 57

\bibitem[{{Fan}, {Ba{\~n}ados} \& {Simcoe}(2023){Fan}, {Ba{\~n}ados}, \& {Simcoe}}]{Fan_2023}
{Fan} X., {Ba{\~n}ados} E., {Simcoe} R.~A., 2023, \araa, 61, 373

\bibitem[{Fernandez {et~al}\mbox{.}(2014)Fernandez, Bryan, Haiman, \& Li}]{fernandez2014}
Fernandez R., Bryan G.~L., Haiman Z., Li M., 2014, Monthly Notices of the Royal Astronomical Society, 439, 3798

\bibitem[{{Fragione} {et~al}\mbox{.}(2022){Fragione}, {Kocsis}, {Rasio}, \& {Silk}}]{Fragione_2022}
{Fragione} G., {Kocsis} B., {Rasio} F.~A., {Silk} J., 2022, \apj, 927, 231

\bibitem[{Fragione \& Leigh(2018)}]{Fragione_2018}
Fragione G., Leigh N., 2018, Monthly Notices of the Royal Astronomical Society, 480, 5160

\bibitem[{Franchini, Sesana \& Dotti(2021)Franchini, Sesana, \& Dotti}]{franchiniCircumbinaryDiscSelfgravity2021}
Franchini A., Sesana A., Dotti M., 2021, Monthly Notices of the Royal Astronomical Society, stab2234

\bibitem[{{Ghez} {et~al}\mbox{.}(2008){Ghez}, {Salim}, {Weinberg}, {Lu}, {Do}, {Dunn}, {Matthews}, {Morris}, {Yelda}, {Becklin}, {Kremenek}, {Milosavljevic}, \& {Naiman}}]{Ghez_2008}
{Ghez} A.~M. {et~al.}, 2008, \apj, 689, 1044

\bibitem[{{Gonz{\'a}lez} {et~al}\mbox{.}(2021){Gonz{\'a}lez}, {Kremer}, {Chatterjee}, {Fragione}, {Rodriguez}, {Weatherford}, {Ye}, \& {Rasio}}]{Gonzalez_2021}
{Gonz{\'a}lez} E., {Kremer} K., {Chatterjee} S., {Fragione} G., {Rodriguez} C.~L., {Weatherford} N.~C., {Ye} C.~S., {Rasio} F.~A., 2021, \apjl, 908, L29

\bibitem[{Gordon {et~al}\mbox{.}(2024)Gordon, Smith, Khochfar, \& Regan}]{gordonHungryNotHow2024}
Gordon S., Smith B., Khochfar S., Regan J., 2024, Hungry or {Not}: {How} {Stellar}-{Mass} {Black} {Holes} {Grow} (or {Don}'t) in {Dark} {Matter} {Mini}-{Haloes} at {High}-{Resolution}. ArXiv:2401.04183 [astro-ph]

\bibitem[{{Goulding} {et~al}\mbox{.}(2023){Goulding}, {Greene}, {Setton}, {Labbe}, {Bezanson}, {Miller}, {Atek}, {Bogdan}, {Brammer}, {Chemerynska}, {Cutler}, {Dayal}, {Fudamoto}, {Fujimoto}, {Furtak}, {Kokorev}, {Khullar}, {Leja}, {Marchesini}, {Natarajan}, {Nelson}, {Oesch}, {Pan}, {Papovich}, {Price}, {van Dokkum}, {Wang}, {Weaver}, {Whitaker}, \& {Zitrin}}]{Goulding_2023}
{Goulding} A.~D. {et~al.}, 2023, arXiv e-prints, arXiv:2308.02750

\bibitem[{{Greene} {et~al}\mbox{.}(2024){Greene}, {Labbe}, {Goulding}, {Furtak}, {Chemerynska}, {Kokorev}, {Dayal}, {Volonteri}, {Williams}, {Wang}, {Setton}, {Burgasser}, {Bezanson}, {Atek}, {Brammer}, {Cutler}, {Feldmann}, {Fujimoto}, {Glazebrook}, {de Graaff}, {Khullar}, {Leja}, {Marchesini}, {Maseda}, {Matthee}, {Miller}, {Naidu}, {Nanayakkara}, {Oesch}, {Pan}, {Papovich}, {Price}, {van Dokkum}, {Weaver}, {Whitaker}, \& {Zitrin}}]{Greene_2024}
{Greene} J.~E. {et~al.}, 2024, \apj, 964, 39

\bibitem[{{Habouzit} {et~al}\mbox{.}(2016){Habouzit}, {Volonteri}, {Latif}, {Dubois}, \& {Peirani}}]{Habouzit_2016}
{Habouzit} M., {Volonteri} M., {Latif} M., {Dubois} Y., {Peirani} S., 2016, \mnras, 463, 529

\bibitem[{Haehnelt(1994)}]{haehneltLowfrequencyGravitationalWaves1994}
Haehnelt M.~G., 1994, Monthly Notices of the Royal Astronomical Society, 269, 199

\bibitem[{{Hartwig} {et~al}\mbox{.}(2015){Hartwig}, {Bromm}, {Klessen}, \& {Glover}}]{Hartwig_2015}
{Hartwig} T., {Bromm} V., {Klessen} R.~S., {Glover} S. C.~O., 2015, \mnras, 447, 3892

\bibitem[{Heath \& Nixon(2020)}]{heathOrbitalEvolutionBinaries2020}
Heath R., Nixon C., 2020, On the orbital evolution of binaries with circumbinary discs. ArXiv:2007.11592 [astro-ph]

\bibitem[{{Henden} {et~al}\mbox{.}(2018){Henden}, {Puchwein}, {Shen}, \& {Sijacki}}]{FABLE}
{Henden} N.~A., {Puchwein} E., {Shen} S., {Sijacki} D., 2018, \mnras, 479, 5385

\bibitem[{{Hosokawa}, {Omukai} \& {Yorke}(2012){Hosokawa}, {Omukai}, \& {Yorke}}]{hosokawaSupermassive2012}
{Hosokawa} T., {Omukai} K., {Yorke} H.~W., 2012, \apj, 756, 93

\bibitem[{{Hoyle} \& {Lyttleton}(1939)}]{1939PCPS...35..405H}
{Hoyle} F., {Lyttleton} R.~A., 1939, Proceedings of the Cambridge Philosophical Society, 35, 405

\bibitem[{Hu \& Wu(2017)}]{Taiji}
Hu W.-R., Wu Y.-L., 2017, Natl. Sci. Rev., 4, 685

\bibitem[{{Inayoshi}, {Visbal} \& {Haiman}(2020){Inayoshi}, {Visbal}, \& {Haiman}}]{Inayoshi_2020}
{Inayoshi} K., {Visbal} E., {Haiman} Z., 2020, ARA\&A, in press; e-print arXiv:1911.05791, arXiv:1911.05791

\bibitem[{{Katz}, {Sijacki} \& {Haehnelt}(2015){Katz}, {Sijacki}, \& {Haehnelt}}]{Katz_2015}
{Katz} H., {Sijacki} D., {Haehnelt} M.~G., 2015, \mnras, 451, 2352

\bibitem[{Kocsis {et~al}\mbox{.}(2006)Kocsis, Frei, Haiman, \& Menou}]{kocsisFindingElectromagneticCounterparts2006}
Kocsis B., Frei Z., Haiman Z., Menou K., 2006, The Astrophysical Journal, 637, 27, arXiv:astro-ph/0505394

\bibitem[{Kormendy \& Ho(2013)}]{doi:10.1146/annurev-astro-082708-101811}
Kormendy J., Ho L.~C., 2013, Annual Review of Astronomy and Astrophysics, 51, 511

\bibitem[{Kulkarni, Visbal \& Bryan(2021)Kulkarni, Visbal, \& Bryan}]{kulkarniCriticalDarkMatter2021}
Kulkarni M., Visbal E., Bryan G.~L., 2021, The Astrophysical Journal, 917, 40, arXiv:2010.04169 [astro-ph]

\bibitem[{{Larson} {et~al}\mbox{.}(2023){Larson}, {Finkelstein}, {Kocevski}, {Hutchison}, {Trump}, {Haro}, {Bromm}, {Cleri}, {Dickinson}, {Fujimoto}, {Kartaltepe}, {Koekemoer}, {Papovich}, {Pirzkal}, {Tacchella}, {Zavala}, {Bagley}, {Behroozi}, {Champagne}, {Cole}, {Jung}, {Morales}, {Yang}, {Zhang}, {Zitrin}, {Amor{\'\i}n}, {Burgarella}, {Casey}, {Ch{\'a}vez Ortiz}, {Cox}, {Chworowsky}, {Fontana}, {Gawiser}, {Grazian}, {Grogin}, {Harish}, {Hathi}, {Hirschmann}, {Holwerda}, {Juneau}, {Leung}, {Lucas}, {McGrath}, {P{\'e}rez-Gonz{\'a}lez}, {Rigby}, {Seill{\'e}}, {Simons}, {de La Vega}, {Weiner}, {Wilkins}, {Yung}, \& {Ceers Team}}]{Larson_2023}
{Larson} R.~L. {et~al.}, 2023, \apjl, 953, L29

\bibitem[{{Latif} {et~al}\mbox{.}(2015){Latif}, {Bovino}, {Grassi}, {Schleicher}, \& {Spaans}}]{Latif_2015}
{Latif} M.~A., {Bovino} S., {Grassi} T., {Schleicher} D.~R.~G., {Spaans} M., 2015, \mnras, 446, 3163

\bibitem[{{Latif}, {Schleicher} \& {Hartwig}(2016){Latif}, {Schleicher}, \& {Hartwig}}]{Latif_2016a}
{Latif} M.~A., {Schleicher} D.~R.~G., {Hartwig} T., 2016, \mnras, 458, 233

\bibitem[{Latif, Schleicher \& Khochfar(2023)Latif, Schleicher, \& Khochfar}]{latifRoleMagneticFields2023}
Latif M.~A., Schleicher D. R.~G., Khochfar S., 2023, The Astrophysical Journal, 945, 137

\bibitem[{{Latif} {et~al}\mbox{.}(2013{\natexlab{a}}){Latif}, {Schleicher}, {Schmidt}, \& {Niemeyer}}]{Latif_2013c}
{Latif} M.~A., {Schleicher} D.~R.~G., {Schmidt} W., {Niemeyer} J., 2013{\natexlab{a}}, \mnras, 433, 1607

\bibitem[{{Latif} {et~al}\mbox{.}(2013{\natexlab{b}}){Latif}, {Schleicher}, {Schmidt}, \& {Niemeyer}}]{Latif_2013d}
{Latif} M.~A., {Schleicher} D.~R.~G., {Schmidt} W., {Niemeyer} J.~C., 2013{\natexlab{b}}, \mnras, 436, 2989

\bibitem[{{Latif} {et~al}\mbox{.}(2022){Latif}, {Whalen}, {Khochfar}, {Herrington}, \& {Woods}}]{Latif_2022}
{Latif} M.~A., {Whalen} D.~J., {Khochfar} S., {Herrington} N.~P., {Woods} T.~E., 2022, \nat, 607, 48

\bibitem[{{Lazar} \& {Bromm}(2022)}]{Lazar_2022}
{Lazar} A., {Bromm} V., 2022, \mnras, 511, 2505

\bibitem[{{Li} {et~al}\mbox{.}(2024){Li}, {Silverman}, {Shen}, {Volonteri}, {Jahnke}, {Zhuang}, {Scoggins}, {Ding}, {Harikane}, {Onoue}, \& {Tanaka}}]{JunyaoTipOfTheIceberg}
{Li} J. {et~al.}, 2024, arXiv e-prints, arXiv:2403.00074

\bibitem[{Li {et~al}\mbox{.}(2024)Li, Silverman, Shen, Volonteri, Jahnke, Zhuang, Scoggins, Ding, Harikane, Onoue, \& Tanaka}]{liTipIcebergOvermassive2024}
Li J. {et~al.}, 2024, Tip of the iceberg: overmassive black holes at 4{\textless}z{\textless}7 found by {JWST} are not inconsistent with the local \${\textbackslash}mathcal\{{M}\}\_\{{\textbackslash}rm {BH}\}\$-\${\textbackslash}mathcal\{{M}\}\_{\textbackslash}star\$ relation. ArXiv:2403.00074 [astro-ph]

\bibitem[{Lodato \& Natarajan(2006)}]{lodatoSupermassiveBlackHole2006}
Lodato G., Natarajan P., 2006, Monthly Notices of the Royal Astronomical Society, 371, 1813

\bibitem[{Lodato \& Natarajan(2007)}]{lodatoMassFunctionHighredshift2007}
Lodato G., Natarajan P., 2007, Monthly Notices of the Royal Astronomical Society: Letters, 377, L64

\bibitem[{Loeb \& Rasio(1994)}]{loebCollapsePrimordialGas1994}
Loeb A., Rasio F.~A., 1994, The Astrophysical Journal, 432, 52

\bibitem[{Luo {et~al}\mbox{.}(2016)Luo, Chen, Duan, Gong, Hu, Ji, Liu, Mei, Milyukov, Sazhin, Shao, Toth, Tu, Wang, Wang, Yeh, Zhan, Zhang, Zharov, \& Zhou}]{Tianqin}
Luo J. {et~al.}, 2016, Classical and Quantum Gravity, 33, 035010

\bibitem[{{Lupi} {et~al}\mbox{.}(2014){Lupi}, {Colpi}, {Devecchi}, {Galanti}, \& {Volonteri}}]{Lupi_2014}
{Lupi} A., {Colpi} M., {Devecchi} B., {Galanti} G., {Volonteri} M., 2014, \mnras, 442, 3616

\bibitem[{{Lupi} {et~al}\mbox{.}(2016){Lupi}, {Haardt}, {Dotti}, {Fiacconi}, {Mayer}, \& {Madau}}]{Lupi_2016}
{Lupi} A., {Haardt} F., {Dotti} M., {Fiacconi} D., {Mayer} L., {Madau} P., 2016, \mnras, 456, 2993

\bibitem[{{Lupi}, {Haiman} \& {Volonteri}(2021){Lupi}, {Haiman}, \& {Volonteri}}]{Lupi_2021}
{Lupi} A., {Haiman} Z., {Volonteri} M., 2021, \mnras, 503, 5046

\bibitem[{{Lupi} {et~al}\mbox{.}(2024){Lupi}, {Trinca}, {Volonteri}, {Dotti}, \& {Mazzucchelli}}]{Lupi_2024}
{Lupi} A., {Trinca} A., {Volonteri} M., {Dotti} M., {Mazzucchelli} C., 2024, arXiv e-prints, arXiv:2406.17847

\bibitem[{{Ma} {et~al}\mbox{.}(2021){Ma}, {Hopkins}, {Ma}, {Angl{\'e}s-Alc{\'a}zar}, {Faucher-Gigu{\`e}re}, \& {Kelley}}]{Ma_2021}
{Ma} L., {Hopkins} P.~F., {Ma} X., {Angl{\'e}s-Alc{\'a}zar} D., {Faucher-Gigu{\`e}re} C.-A., {Kelley} L.~Z., 2021, \mnras, 508, 1973

\bibitem[{{Madau} \& {Rees}(2001)}]{Madau_2001}
{Madau} P., {Rees} M.~J., 2001, \apjl, 551, L27

\bibitem[{Maggiore {et~al}\mbox{.}(2020)Maggiore, Broeck, Bartolo, Belgacem, Bertacca, Bizouard, Branchesi, Clesse, Foffa, García-Bellido, Grimm, Harms, Hinderer, Matarrese, Palomba, Peloso, Ricciardone, \& Sakellariadou}]{EinsteinTelescope}
Maggiore M. {et~al.}, 2020, Journal of Cosmology and Astroparticle Physics, 2020, 050, arXiv:1912.02622 [astro-ph, physics:gr-qc]

\bibitem[{{Maiolino} {et~al}\mbox{.}(2024){Maiolino}, {Scholtz}, {Witstok}, {Carniani}, {D'Eugenio}, {de Graaff}, {{\"U}bler}, {Tacchella}, {Curtis-Lake}, {Arribas}, {Bunker}, {Charlot}, {Chevallard}, {Curti}, {Looser}, {Maseda}, {Rawle}, {Rodr{\'\i}guez del Pino}, {Willott}, {Egami}, {Eisenstein}, {Hainline}, {Robertson}, {Williams}, {Willmer}, {Baker}, {Boyett}, {DeCoursey}, {Fabian}, {Helton}, {Ji}, {Jones}, {Kumari}, {Laporte}, {Nelson}, {Perna}, {Sandles}, {Shivaei}, \& {Sun}}]{Maiolino_2023}
{Maiolino} R. {et~al.}, 2024, \nat, 627, 59

\bibitem[{{Mapelli} {et~al}\mbox{.}(2021){Mapelli}, {Dall'Amico}, {Bouffanais}, {Giacobbo}, {Arca Sedda}, {Artale}, {Ballone}, {Di Carlo}, {Iorio}, {Santoliquido}, \& {Torniamenti}}]{Mapelli_2021}
{Mapelli} M. {et~al.}, 2021, \mnras, 505, 339

\bibitem[{{Matthee} {et~al}\mbox{.}(2023){Matthee}, {Naidu}, {Brammer}, {Chisholm}, {Eilers}, {Goulding}, {Greene}, {Kashino}, {Labbe}, {Lilly}, {Mackenzie}, {Oesch}, {Weibel}, {Wuyts}, {Xiao}, {Bordoloi}, {Bouwens}, {van Dokkum}, {Illingworth}, {Kramarenko}, {Maseda}, {Mason}, {Meyer}, {Nelson}, {Reddy}, {Shivaei}, {Simcoe}, \& {Yue}}]{Matthee_2023}
{Matthee} J. {et~al.}, 2023, arXiv e-prints, arXiv:2306.05448

\bibitem[{{McAlpine} {et~al}\mbox{.}(2016){McAlpine}, {Helly}, {Schaller}, {Trayford}, {Qu}, {Furlong}, {Bower}, {Crain}, {Schaye}, {Theuns}, {Dalla Vecchia}, {Frenk}, {McCarthy}, {Jenkins}, {Rosas-Guevara}, {White}, {Baes}, {Camps}, \& {Lemson}}]{EAGLE}
{McAlpine} S. {et~al.}, 2016, Astronomy and Computing, 15, 72

\bibitem[{{Mehta}, {Regan} \& {Prole}(2024){Mehta}, {Regan}, \& {Prole}}]{Mehta_2004}
{Mehta} D., {Regan} J.~A., {Prole} L., 2024, arXiv e-prints, arXiv:2409.08326

\bibitem[{Mikkola \& Valtonen(1992)}]{mikkolaEvolutionBinariesField1992}
Mikkola S., Valtonen M.~J., 1992, Monthly Notices of the Royal Astronomical Society, 259, 115

\bibitem[{{Milosavljevi{\'c}}, {Couch} \& {Bromm}(2009){Milosavljevi{\'c}}, {Couch}, \& {Bromm}}]{Milosavljevic_2009}
{Milosavljevi{\'c}} M., {Couch} S.~M., {Bromm} V., 2009, \apjl, 696, L146

\bibitem[{{Natarajan}(2020)}]{Natarajan_2020}
{Natarajan} P., 2020, \mnras

\bibitem[{{Natarajan} {et~al}\mbox{.}(2024){Natarajan}, {Pacucci}, {Ricarte}, {Bogd{\'a}n}, {Goulding}, \& {Cappelluti}}]{Natarajan_2024}
{Natarajan} P., {Pacucci} F., {Ricarte} A., {Bogd{\'a}n} {\'A}., {Goulding} A.~D., {Cappelluti} N., 2024, \apjl, 960, L1

\bibitem[{Ni {et~al}\mbox{.}(2022)Ni, Di~Matteo, Chen, Croft, \& Bird}]{niUltramassiveBlackHoles2022}
Ni Y., Di~Matteo T., Chen N., Croft R., Bird S., 2022, The Astrophysical Journal Letters, 940, L49

\bibitem[{{O'Shea} {et~al}\mbox{.}(2015){O'Shea}, {Wise}, {Xu}, \& {Norman}}]{OShea_2015}
{O'Shea} B.~W., {Wise} J.~H., {Xu} H., {Norman} M.~L., 2015, \apjl, 807, L12

\bibitem[{{Pacucci}, {Mezcua} \& {Regan}(2021){Pacucci}, {Mezcua}, \& {Regan}}]{Pacucci_2021}
{Pacucci} F., {Mezcua} M., {Regan} J.~A., 2021, \apj, 920, 134

\bibitem[{Pacucci {et~al}\mbox{.}(2023)Pacucci, Nguyen, Carniani, Maiolino, \& Fan}]{pacucci2023jwst}
Pacucci F., Nguyen B., Carniani S., Maiolino R., Fan X., 2023, Jwst ceers and jades active galaxies at z = 4-7 violate the local $m_\bullet-m_\star$ relation at $>3\sigma$: Implications for low-mass black holes and seeding models

\bibitem[{Peters \& Mathews(1963)}]{petersGravitationalRadiationPoint1963}
Peters P.~C., Mathews J., 1963, Physical Review, 131, 435

\bibitem[{Peters {et~al}\mbox{.}(2014)Peters, Schleicher, Smith, Schmidt, \& Klessen}]{petersLowmetallicityStarFormation2014}
Peters T., Schleicher D. R.~G., Smith R.~J., Schmidt W., Klessen R.~S., 2014, Monthly Notices of the Royal Astronomical Society, 442, 3112

\bibitem[{Pfister {et~al}\mbox{.}(2019)Pfister, Volonteri, Dubois, Dotti, \& Colpi}]{pfisterErraticDynamicalLife2019}
Pfister H., Volonteri M., Dubois Y., Dotti M., Colpi M., 2019, Monthly Notices of the Royal Astronomical Society, 486, 101, arXiv:1902.01297 [astro-ph]

\bibitem[{{Prole} {et~al}\mbox{.}(2024){Prole}, {Regan}, {Glover}, {Klessen}, {Priestley}, \& {Clark}}]{Prole_2024}
{Prole} L.~R., {Regan} J.~A., {Glover} S. C.~O., {Klessen} R.~S., {Priestley} F.~D., {Clark} P.~C., 2024, \aap, 685, A31

\bibitem[{Quinlan(1996)}]{quinlanDynamicalEvolutionMassive1996}
Quinlan G.~D., 1996, New Astronomy, 1, 35

\bibitem[{Quinlan \& Shapiro(1987)}]{quinlanCollapseDenseStar1987}
Quinlan G.~D., Shapiro S.~L., 1987, The Astrophysical Journal, 321, 199

\bibitem[{{Rees}(1978)}]{Rees_1978}
{Rees} M.~J., 1978, Physica Scripta, 17, 193

\bibitem[{{Regan}(2023)}]{Regan_2023}
{Regan} J., 2023, The Open Journal of Astrophysics, 6, 12

\bibitem[{{Regan} \& {Volonteri}(2024)}]{Regan_2024}
{Regan} J., {Volonteri} M., 2024, arXiv e-prints, arXiv:2405.17975

\bibitem[{{Regan} \& {Downes}(2018)}]{Regan_2018b}
{Regan} J.~A., {Downes} T.~P., 2018, \mnras, 478, 5037

\bibitem[{{Regan} {et~al}\mbox{.}(2020){Regan}, {Haiman}, {Wise}, {O'Shea}, \& {Norman}}]{Regan_2020a}
{Regan} J.~A., {Haiman} Z., {Wise} J.~H., {O'Shea} B.~W., {Norman} M.~L., 2020, The Open Journal of Astrophysics, 3, E9

\bibitem[{{Regan}, {Johansson} \& {Wise}(2014){Regan}, {Johansson}, \& {Wise}}]{Regan_2014}
{Regan} J.~A., {Johansson} P.~H., {Wise} J.~H., 2014, \apj, 795, 137

\bibitem[{{Regan}, {Johansson} \& {Wise}(2016){Regan}, {Johansson}, \& {Wise}}]{Regan_2016}
{Regan} J.~A., {Johansson} P.~H., {Wise} J.~H., 2016, \mnras, 459, 3377

\bibitem[{{Regan} {et~al}\mbox{.}(2017){Regan}, {Visbal}, {Wise}, {Haiman}, {Johansson}, \& {Bryan}}]{Regan_2017}
{Regan} J.~A., {Visbal} E., {Wise} J.~H., {Haiman} Z., {Johansson} P.~H., {Bryan} G.~L., 2017, Nature Astronomy, 1, 0075

\bibitem[{Regan {et~al}\mbox{.}(2020)Regan, Wise, O'Shea, \& Norman}]{reganEmergenceFirstStarfree2020}
Regan J.~A., Wise J.~H., O'Shea B.~W., Norman M.~L., 2020, Monthly Notices of the Royal Astronomical Society, 492, 3021, arXiv:1908.02823 [astro-ph]

\bibitem[{{Regan} {et~al}\mbox{.}(2020){Regan}, {Wise}, {Woods}, {Downes}, {O'Shea}, \& {Norman}}]{Regan_2020b}
{Regan} J.~A., {Wise} J.~H., {Woods} T.~E., {Downes} T.~P., {O'Shea} B.~W., {Norman} M.~L., 2020, The Open Journal of Astrophysics, 3, 15

\bibitem[{Regan {et~al}\mbox{.}(2020)Regan, Wise, Woods, Downes, O’Shea, \& Norman}]{Regan_2020}
Regan J.~A., Wise J.~H., Woods T.~E., Downes T.~P., O’Shea B.~W., Norman M.~C., 2020, The Open Journal of Astrophysics, 3

\bibitem[{Reines \& Volonteri(2015)}]{reinesRelationsCentralBlack2015}
Reines A.~E., Volonteri M., 2015, The Astrophysical Journal, 813, 82, arXiv:1508.06274 [astro-ph]

\bibitem[{Reinoso {et~al}\mbox{.}(2023)Reinoso, Klessen, Schleicher, Glover, \& Solar}]{reinosoFormationSupermassiveStars2023}
Reinoso B., Klessen R.~S., Schleicher D., Glover S. C.~O., Solar P., 2023, Monthly Notices of the Royal Astronomical Society, 521, 3553

\bibitem[{{Reitze} {et~al}\mbox{.}(2019){Reitze}, {Adhikari}, {Ballmer}, {Barish}, {Barsotti}, {Billingsley}, {Brown}, {Chen}, {Coyne}, {Eisenstein}, {Evans}, {Fritschel}, {Hall}, {Lazzarini}, {Lovelace}, {Read}, {Sathyaprakash}, {Shoemaker}, {Smith}, {Torrie}, {Vitale}, {Weiss}, {Wipf}, \& {Zucker}}]{Cosmic_explorer}
{Reitze} D. {et~al.}, 2019, in Bulletin of the American Astronomical Society, Vol.~51, p.~35

\bibitem[{{Rizzuto} {et~al}\mbox{.}(2021){Rizzuto}, {Naab}, {Spurzem}, {Giersz}, {Ostriker}, {Stone}, {Wang}, {Berczik}, \& {Rampp}}]{Rizzuto_2021}
{Rizzuto} F.~P. {et~al.}, 2021, \mnras, 501, 5257

\bibitem[{Robson, Cornish \& Liu(2019)Robson, Cornish, \& Liu}]{Robson_2019}
Robson T., Cornish N.~J., Liu C., 2019, Classical and Quantum Gravity, 36, 105011

\bibitem[{{Saglia} {et~al}\mbox{.}(2016){Saglia}, {Opitsch}, {Erwin}, {Thomas}, {Beifiori}, {Fabricius}, {Mazzalay}, {Nowak}, {Rusli}, \& {Bender}}]{2016ApJ...818...47S}
{Saglia} R.~P. {et~al.}, 2016, \apj, 818, 47

\bibitem[{Saini, Bhat \& Arun(2022)Saini, Bhat, \& Arun}]{sainiPremergerLocalizationIntermediate2022}
Saini P., Bhat S.~A., Arun K.~G., 2022, Physical Review D, 106, 104015, arXiv:2208.03004 [astro-ph, physics:gr-qc]

\bibitem[{Sassano {et~al}\mbox{.}(2021)Sassano, Schneider, Valiante, Inayoshi, Chon, Omukai, Mayer, \& Capelo}]{sassanoLightMediumweightHeavy2021}
Sassano F., Schneider R., Valiante R., Inayoshi K., Chon S., Omukai K., Mayer L., Capelo P.~R., 2021, Monthly Notices of the Royal Astronomical Society, 506, 613, arXiv:2106.08330 [astro-ph]

\bibitem[{Schauer {et~al}\mbox{.}(2017)Schauer, Regan, Glover, \& Klessen}]{schauerFormationDirectCollapse2017}
Schauer A. T.~P., Regan J., Glover S. C.~O., Klessen R.~S., 2017, Monthly Notices of the Royal Astronomical Society, 471, 4878, arXiv:1705.02347 [astro-ph]

\bibitem[{Schaye {et~al}\mbox{.}(2015)Schaye, Crain, Bower, Furlong, Schaller, Theuns, Dalla~Vecchia, Frenk, McCarthy, Helly, Jenkins, Rosas-Guevara, White, Baes, Booth, Camps, Navarro, Qu, Rahmati, Sawala, Thomas, \& Trayford}]{schayeEAGLEProjectSimulating2015}
Schaye J. {et~al.}, 2015, Monthly Notices of the Royal Astronomical Society, 446, 521

\bibitem[{Schleicher {et~al}\mbox{.}(2022)Schleicher, Reinoso, Latif, Klessen, Vergara, Das, Alister, Díaz, \& Solar}]{schleicherOriginSupermassiveBlack2022}
Schleicher D. R.~G. {et~al.}, 2022, Monthly Notices of the Royal Astronomical Society, 512, 6192

\bibitem[{{Schneider} {et~al}\mbox{.}(2002){Schneider}, {Ferrara}, {Natarajan}, \& {Omukai}}]{Schnedier_2002}
{Schneider} R., {Ferrara} A., {Natarajan} P., {Omukai} K., 2002, \apj, 571, 30

\bibitem[{{Scholtz} {et~al}\mbox{.}(2023){Scholtz}, {Maiolino}, {D'Eugenio}, {Curtis-Lake}, {Carniani}, {Charlot}, {Curti}, {Silcock}, {Arribas}, {Baker}, {Bhatawdekar}, {Boyett}, {Bunker}, {Chevallard}, {Circosta}, {Eisenstein}, {Hainline}, {Hausen}, {Ji}, {Ji}, {Johnson}, {Kumari}, {Looser}, {Lyu}, {Maseda}, {Parlanti}, {Perna}, {Rieke}, {Robertson}, {Rodr{\'\i}guez Del Pino}, {Sun}, {Tacchella}, {{\"U}bler}, {Venturi}, {Williams}, {Willmer}, {Willott}, \& {Witstok}}]{Scholtz_2023}
{Scholtz} J. {et~al.}, 2023, arXiv e-prints, arXiv:2311.18731

\bibitem[{{Scoggins}, {Haiman} \& {Wise}(2023){Scoggins}, {Haiman}, \& {Wise}}]{Scoggins2023}
{Scoggins} M.~T., {Haiman} Z., {Wise} J.~H., 2023, \mnras, 519, 2155

\bibitem[{Sesana, Haardt \& Madau(2006)Sesana, Haardt, \& Madau}]{sesanaInteractionMassiveBlack2006}
Sesana A., Haardt F., Madau P., 2006, The Astrophysical Journal, 651, 392, arXiv:astro-ph/0604299

\bibitem[{Sesana, Volonteri \& Haardt(2007)Sesana, Volonteri, \& Haardt}]{sesanaImprintMassiveBlack2007}
Sesana A., Volonteri M., Haardt F., 2007, Monthly Notices of the Royal Astronomical Society, 377, 1711

\bibitem[{{Sesana}, {Volonteri} \& {Haardt}(2007){Sesana}, {Volonteri}, \& {Haardt}}]{Sesana_2007}
{Sesana} A., {Volonteri} M., {Haardt} F., 2007, \mnras, 377, 1711

\bibitem[{{Sesana}, {Volonteri} \& {Haardt}(2009){Sesana}, {Volonteri}, \& {Haardt}}]{Sesana_2009}
{Sesana} A., {Volonteri} M., {Haardt} F., 2009, Classical and Quantum Gravity, 26, 094033

\bibitem[{{Shang}, {Bryan} \& {Haiman}(2010){Shang}, {Bryan}, \& {Haiman}}]{Shang_2010}
{Shang} C., {Bryan} G.~L., {Haiman} Z., 2010, \mnras, 402, 1249

\bibitem[{{Shi} {et~al}\mbox{.}(2023){Shi}, {Kremer}, {Grudi{\'c}}, {Gerling-Dunsmore}, \& {Hopkins}}]{Shi_2023}
{Shi} Y., {Kremer} K., {Grudi{\'c}} M.~Y., {Gerling-Dunsmore} H.~J., {Hopkins} P.~F., 2023, \mnras, 518, 3606

\bibitem[{{Shi}, {Kremer} \& {Hopkins}(2024){Shi}, {Kremer}, \& {Hopkins}}]{Shi_2024}
{Shi} Y., {Kremer} K., {Hopkins} P.~F., 2024, arXiv e-prints, arXiv:2405.12164

\bibitem[{Shi, Kremer \& Hopkins(2024)Shi, Kremer, \& Hopkins}]{Shi_2024b}
Shi Y., Kremer K., Hopkins P.~F., 2024

\bibitem[{Sijacki {et~al}\mbox{.}(2015)Sijacki, Vogelsberger, Genel, Springel, Torrey, Snyder, Nelson, \& Hernquist}]{sijackiIllustrisSimulationEvolving2015}
Sijacki D., Vogelsberger M., Genel S., Springel V., Torrey P., Snyder G.~F., Nelson D., Hernquist L., 2015, Monthly Notices of the Royal Astronomical Society, 452, 575

\bibitem[{Smith \& Lang(2019)}]{ytree}
Smith B.~D., Lang M., 2019, Journal of Open Source Software, 4, 1881

\bibitem[{Smith {et~al}\mbox{.}(2018{\natexlab{a}})Smith, Regan, Downes, Norman, O'Shea, \& Wise}]{smithGrowthBlackHoles2018}
Smith B.~D., Regan J.~A., Downes T.~P., Norman M.~L., O'Shea B.~W., Wise J.~H., 2018{\natexlab{a}}, Monthly Notices of the Royal Astronomical Society, 480, 3762, aDS Bibcode: 2018MNRAS.480.3762S

\bibitem[{Smith {et~al}\mbox{.}(2018{\natexlab{b}})Smith, Regan, Downes, Norman, O'Shea, \& Wise}]{Smith_2018}
Smith B.~D., Regan J.~A., Downes T.~P., Norman M.~L., O'Shea B.~W., Wise J.~H., 2018{\natexlab{b}}, Monthly Notices of the Royal Astronomical Society, 480, 3762

\bibitem[{{Spinoso} {et~al}\mbox{.}(2023){Spinoso}, {Bonoli}, {Valiante}, {Schneider}, \& {Izquierdo-Villalba}}]{Spinoso_2024}
{Spinoso} D., {Bonoli} S., {Valiante} R., {Schneider} R., {Izquierdo-Villalba} D., 2023, \mnras, 518, 4672

\bibitem[{Spolyar, Freese \& Gondolo(2008)Spolyar, Freese, \& Gondolo}]{DMSTARS-SPOLYAR}
Spolyar D., Freese K., Gondolo P., 2008, Physical Review Letters, 100

\bibitem[{Springel, Di~Matteo \& Hernquist(2005)Springel, Di~Matteo, \& Hernquist}]{springelModellingFeedbackStars2005}
Springel V., Di~Matteo T., Hernquist L., 2005, Monthly Notices of the Royal Astronomical Society, 361, 776

\bibitem[{Tagawa, Haiman \& Kocsis(2020)Tagawa, Haiman, \& Kocsis}]{Tagawa_Haiman_Kocsis_2020}
Tagawa H., Haiman Z., Kocsis B., 2020, The Astrophysical Journal, 892, 36

\bibitem[{Tanaka \& Haiman(2009)}]{tanakaASSEMBLYSUPERMASSIVEBLACK2009}
Tanaka T., Haiman Z., 2009, The Astrophysical Journal, 696, 1798

\bibitem[{Trebitsch {et~al}\mbox{.}(2022)Trebitsch, Hutter, Dayal, Gottlöber, Legrand, \& Yepes}]{trebitschAstraeusVIHierarchical2022}
Trebitsch M., Hutter A., Dayal P., Gottlöber S., Legrand L., Yepes G., 2022, Monthly Notices of the Royal Astronomical Society, 518, 3576

\bibitem[{{Tremmel} {et~al}\mbox{.}(2015){Tremmel}, {Governato}, {Volonteri}, \& {Quinn}}]{Tremmel_2015}
{Tremmel} M., {Governato} F., {Volonteri} M., {Quinn} T.~R., 2015, \mnras, 451, 1868

\bibitem[{{Trinca} {et~al}\mbox{.}(2022){Trinca}, {Schneider}, {Valiante}, {Graziani}, {Zappacosta}, \& {Shankar}}]{Trinca_2022}
{Trinca} A., {Schneider} R., {Valiante} R., {Graziani} L., {Zappacosta} L., {Shankar} F., 2022, \mnras, 511, 616

\bibitem[{Valiante {et~al}\mbox{.}(2020)Valiante, Colpi, Schneider, Mangiagli, Bonetti, Cerini, Fairhurst, Haardt, Mills, \& Sesana}]{valianteUnveilingEarlyBlack2020}
Valiante R. {et~al.}, 2020, Monthly Notices of the Royal Astronomical Society, 500, 4095

\bibitem[{{Valiante} {et~al}\mbox{.}(2021){Valiante}, {Colpi}, {Schneider}, {Mangiagli}, {Bonetti}, {Cerini}, {Fairhurst}, {Haardt}, {Mills}, \& {Sesana}}]{Valiante_2021}
{Valiante} R. {et~al.}, 2021, \mnras, 500, 4095

\bibitem[{{van den Bosch}(2016)}]{2016ApJ...831..134V}
{van den Bosch} R. C.~E., 2016, \apj, 831, 134

\bibitem[{Vergara {et~al}\mbox{.}(2023)Vergara, Escala, Schleicher, \& Reinoso}]{Vergara_Escala_Schleicher_Reinoso_2023}
Vergara M.~C., Escala A., Schleicher D. R.~G., Reinoso B., 2023, Monthly Notices of the Royal Astronomical Society, 522, 4224–4237

\bibitem[{{Visbal}, {Haiman} \& {Bryan}(2014{\natexlab{a}}){Visbal}, {Haiman}, \& {Bryan}}]{Visbal_2014a}
{Visbal} E., {Haiman} Z., {Bryan} G.~L., 2014{\natexlab{a}}, \mnras, 442, L100

\bibitem[{{Visbal}, {Haiman} \& {Bryan}(2014{\natexlab{b}}){Visbal}, {Haiman}, \& {Bryan}}]{Visbal_2014}
{Visbal} E., {Haiman} Z., {Bryan} G.~L., 2014{\natexlab{b}}, \mnras, 445, 1056

\bibitem[{Volonteri(2010)}]{Volonteri_2010}
Volonteri M., 2010, The Astronomy and Astrophysics Review, 18, 279–315

\bibitem[{{Whalen}, {Abel} \& {Norman}(2004){Whalen}, {Abel}, \& {Norman}}]{Whalen_2004}
{Whalen} D., {Abel} T., {Norman} M.~L., 2004, \apj, 610, 14

\bibitem[{{Wise} {et~al}\mbox{.}(2019){Wise}, {Regan}, {O'Shea}, {Norman}, {Downes}, \& {Xu}}]{Wise_2019}
{Wise} J.~H., {Regan} J.~A., {O'Shea} B.~W., {Norman} M.~L., {Downes} T.~P., {Xu} H., 2019, \nat, 566, 85

\bibitem[{Wise {et~al}\mbox{.}(2019)Wise, Regan, O'Shea, Norman, Downes, \& Xu}]{wiseFormationMassiveBlack2019}
Wise J.~H., Regan J.~A., O'Shea B.~W., Norman M.~L., Downes T.~P., Xu H., 2019, Nature, 566, 85, aDS Bibcode: 2019Natur.566...85W

\bibitem[{{Woods} {et~al}\mbox{.}(2019){Woods}, {Agarwal}, {Bromm}, {Bunker}, {Chen}, {Chon}, {Ferrara}, {Glover}, {Haemmerl{\'e}}, {Haiman}, {Hartwig}, {Heger}, {Hirano}, {Hosokawa}, {Inayoshi}, {Klessen}, {Kobayashi}, {Koliopanos}, {Latif}, {Li}, {Mayer}, {Mezcua}, {Natarajan}, {Pacucci}, {Rees}, {Regan}, {Sakurai}, {Salvadori}, {Schneider}, {Surace}, {Tanaka}, {Whalen}, \& {Yoshida}}]{Woods_2019}
{Woods} T.~E. {et~al.}, 2019, \pasa, 36, e027

\bibitem[{Xu {et~al}\mbox{.}(2014)Xu, Ahn, Wise, Norman, \& O'Shea}]{xuHEATINGINTERGALACTICMEDIUM2014a}
Xu H., Ahn K., Wise J.~H., Norman M.~L., O'Shea B.~W., 2014, The Astrophysical Journal, 791, 110

\bibitem[{Xu, Wise \& Norman(2013)Xu, Wise, \& Norman}]{xuPOPULATIONIIISTARS2013a}
Xu H., Wise J.~H., Norman M.~L., 2013, The Astrophysical Journal, 773, 83

\bibitem[{{Yoshida} {et~al}\mbox{.}(2003){Yoshida}, {Abel}, {Hernquist}, \& {Sugiyama}}]{Yoshida_2003}
{Yoshida} N., {Abel} T., {Hernquist} L., {Sugiyama} N., 2003, \apj, 592, 645

\bibitem[{Yungelson {et~al}\mbox{.}(2008)Yungelson, Van Den~Heuvel, Vink, Portegies~Zwart, \& De~Koter}]{Yungelson_VanDenHeuvel_Vink_PortegiesZwart_DeKoter_2008}
Yungelson L.~R., Van Den~Heuvel E. P.~J., Vink J.~S., Portegies~Zwart S.~F., De~Koter A., 2008, Astronomy \& Astrophysics, 477, 223–237

\end{thebibliography}
\end{document}